\documentclass[fleqn,usenatbib]{mnras}

\usepackage{newtxtext,newtxmath}

\usepackage[T1]{fontenc}

\DeclareRobustCommand{\VAN}[3]{#2}
\let\VANthebibliography\thebibliography
\def\thebibliography{\DeclareRobustCommand{\VAN}[3]{##3}\VANthebibliography}

\newcommand{\ag}{{ag}} 
\newcommand{\jgg}{{jgg}} 
\newcommand{\aig}{{aig}} 

\usepackage{graphicx}	
\usepackage{amsmath}	
\DeclareUnicodeCharacter{2212}{-}

\title{Distribution and Recovery Phase of Geomagnetic Storms During Solar Cycles 23 and 24}

\author[W. Mishra et al.]{
Wageesh Mishra \orcid{0000-0003-2740-2280},$^{1}$\thanks{E-mail: wageesh.mishra@iiap.res.in (WM)}
Preity Sukla Sahani \orcid{0009-0008-5121-8524},$^{2}$ 
Soumyaranjan Khuntia \orcid{0009-0006-3209-658X}$^{1, 3}$ and
Dibyendu Chakrabarty \orcid{0000-0003-2693-5325}$^{4}$ 
\\
$^{1}$Indian Institute of Astrophysics, II Block, Koramangala, Bengaluru 560034, India\\
$^{2}$Indian Institute of Science Education and Research, Berhampur 760010, India\\
$^{3}$Pondicherry University, R.V. Nagar, Kalapet 605014, Puducherry, India \\
$^{4}$Space and Atmospheric Sciences Division, Physical Research Laboratory, Ahmedabad, India
}

\pubyear{2023}

\begin{document}
\label{firstpage}
\pagerange{\pageref{firstpage}--\pageref{lastpage}}
\maketitle

\begin{abstract}
Coronal mass ejections (CMEs) and Stream Interaction Regions (SIRs) are the main drivers of intense geomagnetic storms. We study the distribution of geomagnetic storms associated with different drivers during solar cycles 23 and 24 (1996-2019). Although the annual occurrence rate of geomagnetic storms in both cycles tracks the sunspot cycle, the second peak in storm activity lags the second sunspot peak. SIRs contribute significantly to the second peak in storm numbers in both cycles, particularly for moderate to stronger-than-moderate storms. We note semiannual peaks in storm numbers much closer to equinoxes for moderate storms, and slightly shifted from equinoxes for intense and stronger-than-intense storms. We note a significant fraction of multiple-peak storms in both cycles due to isolated ICMEs/SIRs, while single-peak storms from multiple interacting drivers, suggesting a complex relationship between storm steps and their drivers. Our study focuses on investigating the recovery phases of geomagnetic storms and examining their dependencies on various storm parameters. Multiple-peak storms in both cycles have recovery phase duration strongly influenced by slow and fast decay phases with no correlation with the main phase buildup rate and Dst peak. However, the recovery phase in single-peak storms for both cycles depends to some extent on the main phase buildup rate and Dst peak, in addition to slow and fast decay phases. Future research should explore recovery phases of single and multiple-peak storms incorporating in-situ solar wind observations for a deeper understanding of storm evolution and decay processes.
\end{abstract}


\section{Introduction}

 Various energetic phenomena on the Sun result in short- and long-term fluctuations in our geospace \citep{Schwenn2006,Baker2009}. One of these fluctuations is geomagnetic storms, which are short-term disturbances in the Earth's magnetic field \citep{Dessler1960,Gonzalez1994}. Geomagnetic storm intensity is often represented by the disturbance storm time (Dst) index, which measures the perturbation in the horizontal component of the geomagnetic field at equatorial latitudes \citep{Sugiura1964,Iyemori1990,}. A typical geomagnetic storm consists of three phases: initial, main, and recovery. The initial phase is characterized by a small increase in Dst over tens of minutes, typically called a sudden storm commencement (SSC). A substantial decrease in the Dst index over a few hours defines the main phase of a geomagnetic storm. The recovery phase is represented by a slow variation in Dst from its decreased minimum value to its pre-storm level over a few hours to a few days \citep{Gonzalez1999,Echer2008}. An SSC does not need to be present for a storm to occur; hence, the initial phase is not an essential feature of a geomagnetic storm.

 Researchers have previously studied the development and evolution of geomagnetic storms and have established that coronal mass ejections (CMEs), which are large ejections of materials from the Sun into the heliosphere, are the primary drivers of intense geomagnetic storms \citep{Echer2008,Kilpua2017a}. Considering this, several methods have been developed and implemented on CME observations to estimate their arrival time at the Earth \citep{Schwenn2005,Rouillard2008,Davies2013,Mishra2013}. In addition to CMEs, stream interaction regions (SIRs), another type of large-scale solar wind structure, can cause geomagnetic storms \citep{Gosling1993a,Wimmer-Schweingruber1997}. SIRs are formed due to the interaction between high-speed solar wind streams from coronal holes and slow solar wind streams propagating in the interplanetary medium. When SIRs last for one or more solar rotations, they are frequently referred to as co-rotating interaction regions (CIRs) \citep{Kilpua2017,Richardson2018}. It is also possible that the interaction of CMEs with other CMEs or SIRs also drives the geomagnetic storms \citep{Zhang2007}.

CMEs are called interplanetary coronal mass ejections (ICMEs) while traveling through the heliosphere \citep{Forsyth2006,Richardson2010}. It is known that when CMEs and SIR-driven shocks reach Earth, they can compress the geomagnetic field lines at the magnetopause, leading to an increase in the Dst index, termed SSC. When ICMEs and SIRs reach Earth, they can transfer energy to the planet's magnetosphere, causing disruptions to the geomagnetic field \citep{Kilpua2017}. The energy transfer occurs most effectively when the ICMEs/SIRs possess a southward magnetic field component. This southward magnetic field interacts with the Earth's northward magnetic field on the dayside, allowing energy to be deposited into the magnetosphere, creating a westward ring current \citep{Dungey1961,Daglis1999,Tsurutani2006}. The westward magnetospheric ring current generates a magnetic field in the opposite direction to that of the Earth's magnetic field, resulting in a decrease in the Earth's net magnetic field, as reflected in the Dst index \citep{Daglis1999,Kozyra2003}. The final phase of the geomagnetic storm is related to the decay of the ring current, which returns the Earth's magnetic field to a quiet state \citep{Akasofu1963,Akasofu2018}. The growth and decay of the Dst index depend on the relative magnitude of the buildup and decay rate of the ring current. The ring current decay occurs when the injection of particles is no longer strong enough to overcome the loss processes.

Several complex loss mechanisms for ring current exist, such as bounce loss, drift loss, charge exchange with neutrals, coulomb collisions, resonant interactions with electromagnetic ion cyclotron (EMIC) waves, scattering by whistler-mode waves, etc. Also, the ring current gets intervened by three other major current systems: ionospheric currents, field-aligned currents, and magnetotail currents, which contribute differently to the decay of ring current \citep{Campbell1996,Daglis1999}. The composition of ring current is also important as the charge exchange period of $\text{O}^{+}$ ions is less compared to that of $\text{H}^{+}$ ions. Studies have suggested that SIRs are embedded with large amplitude Alfven waves, and the storms driven by them are usually of more extended duration recovery phase \cite{Richardson2018,Telloni2021}. Also, the intensity of such SIRs-driven storms is more under-predicted using models \citep{Liemohn2008}. Earlier studies have established that fast ICMEs arriving at Earth are the dominant drivers of geomagnetic storms around solar maximum. However, during solar minimum, SIRs dominate the source of geomagnetic disturbances \citep{Gonzalez1999}. Therefore, it is crucial to consider several geomagnetic storms driven by different types of solar wind drivers during different phases of solar cycles and understand the recovery phase of these storms.

The 11-year periodic variation in sunspot numbers, known as the sunspot cycle, is known to be well-correlated with the occurrence rate of CMEs from the Sun and ICMEs near the Earth \citep{Webb1994,Wang2014a,Kilpua2014,Lamy2017}. The rate of SIR formation and their arrival at Earth depends on the coronal hole's number, sizes, and heliographic locations, which vary with the solar cycle \citep{Wimmer-Schweingruber1997,Richardson2018}. Coronal holes are more pronounced near solar minimum at higher polar latitudes, but they can grow and migrate to lower solar latitudes around solar maximum. Consequently, the rate of ICMEs and SIRs on Earth and the frequency of storms they drive are expected to depend on different solar cycle phases. It is found that during solar minimum, the Earth is embedded in CIRs for around five times more than ICME structures. However, during solar maximum, the Earth is almost equally embedded in the CMEs and CIRs \citep{Richardson2000}.

An individual ICME and CIR often lead to a one-step classical geomagnetic storm. A classical geomagnetic storm undergoes a ``main phase", eventually reaching a minimum of the Dst index, and then it recovers to pre-storm levels. However, it is noticed that a significant fraction of storms show a more complicated evolution of a two-step decrease in the Dst index during the main phase. It is established that a combination of ICME and SIR structures, such as ICME-ICME or ICME-CIR interacting structures, often give rise to a multi-step, enhanced geomagnetic storm \citep{Burlaga1987,Gonzalez1999,Zhang2007}. The chances of interaction of ICME with another ICME or SIR increase during the maxima of solar cycles. There have been several studies on the behavior of CME-CME interaction and CME-SIR interaction in the interplanetary medium \citep{Gopalswamy2001c,Harrison2012,Mishra2015,Mishra2017,Mishra2021,Palmerio2022}. Moreover, all the storms, irrespective of their drivers, show a two-step recovery phase: a faster recovery and a slower recovery phase \citep{Gonzalez1994,Tsurutani2006}. Since the merged and interacting structures lead to a non-typical development of storms \citep{Gonzalez1999,Farrugia2006,Lugaz2017}, investigating the recovery phase of such storms in contrast to the recovery phase of a classical storm is an interesting avenue to pursue.

Most of the earlier studies have considered the ability of different solar wind drivers, such as ICMEs, SIRs, and plasma parameters therein, to lead to the minimum value of the Dst index \citep{Gonzalez1994,Tsurutani2009,Srivastava2004}. Although the peak of the geomagnetic storm is important to know the severity of the disturbance it can cause, it is also essential to understand how long the magnetosphere remains disturbed before returning to a quiet state. However, only limited studies explore the recovery phase of geomagnetic storms. These handfuls of studies addressing the recovery phase have analyzed either individual cases of storms or only selected intense geomagnetic storms \citep{Liemohn1999,Aguado2010,Yermolaev2012,Cid2013,Raghav2019,Telloni2021}. Thus, it is necessary to investigate the recovery phase of storms corresponding to their different solar wind drivers and over extended periods of solar activity cycles.

We aim to focus our study on the solar cycles 23 and 24 as these two solar cycles differed in the rate of CMEs, ICMEs, and SIRs with sunspot numbers \citep{Chi2018,Richardson2018,Mishra2019}. Also, the average radial sizes of ICMEs and MCs at 1 AU are found to be different in cycles 23 and 24 \citep{Gopalswamy2020,Mishra2021a}. Our study attempts to investigate the distribution of storms of different intensities led by ICMEs, SIRs, and interacting structures over the last two solar cycles. We focus on understanding the recovery phase of the one-step and multi-step geomagnetic storms driven by isolated ICMEs/SIRs and interacting ICME-ICME or ICME-SIR structures. We attempt to estimate the recovery time for the storms and examine their dependency on the main phase buildup rate, main phase duration, peak of the main phase, and slower or faster recovery phase. The availability of continuous observations of the near-Earth solar wind and interplanetary magnetic field as well the Dst index in the last two solar cycles adds an advantage to our study.

\section{Analysis of Distribution of Geomagnetic Storms}

We take the 1-hr Dst data sets from the OMNI Database at the National Space Science Data Center of NASA \citep{Papi2005}, for which the source is the World Data Center for Geomagnetism, Kyoto, \url{https://wdc.kugi.kyoto-u.ac.jp/wdc/Sec3.html}. We used the Dst index to examine the frequency of moderate and stronger-than-moderate storms in both solar cycles 23 and 24. The classification of storms is done as described by \citet{Gonzalez2011} based on the peak Dst value in the main phase. It is as follows: weak geomagnetic storms ($-50\;\text{nT} < \text{Dst} \leq 0\;\text{nT}$), moderate geomagnetic storms ($ -100\;\text{nT} < \text{Dst} \leq -50\;\text{nT}$), intense geomagnetic storms ($-200\;\text{nT} < \text{Dst} \leq -100\;\text{nT}$), severe geomagnetic storms ($-350 \;\text{nT} < \text{Dst} \leq -200\;\text{nT}$), and great geomagnetic storms ($\text{Dst} \leq -350\;\text{nT}$). We compare the storm's occurrence rate with the sunspot numbers, a standard proxy of solar activity. We use a 13-month smoothed monthly sunspot number from ROB's Solar Influences Data Analysis Center (SIDC).

\begin{figure*}

  \centering
   \includegraphics[width=0.45\textwidth]{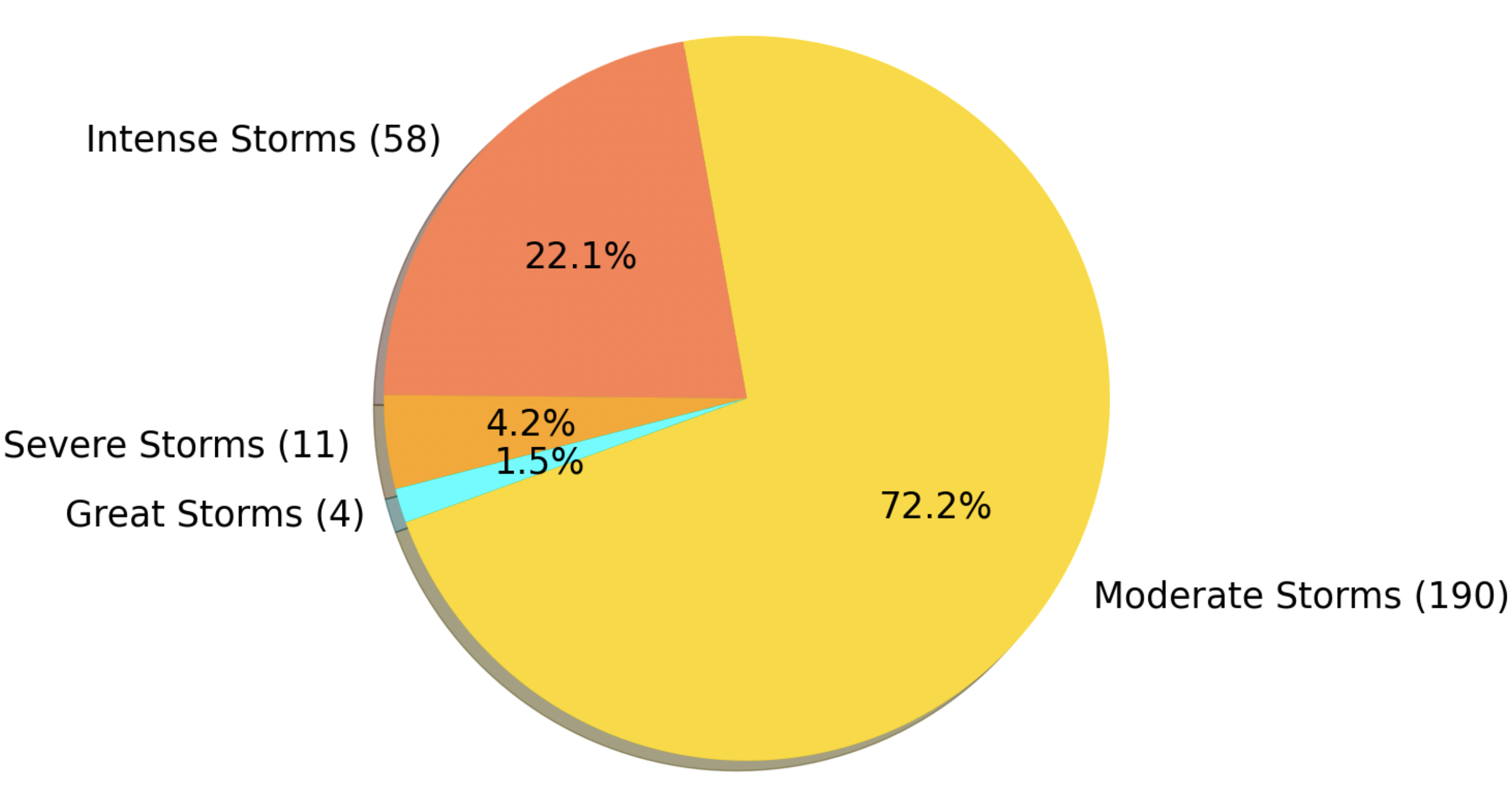}
   \hspace{0.5cm}
   \includegraphics[width=0.45\textwidth]{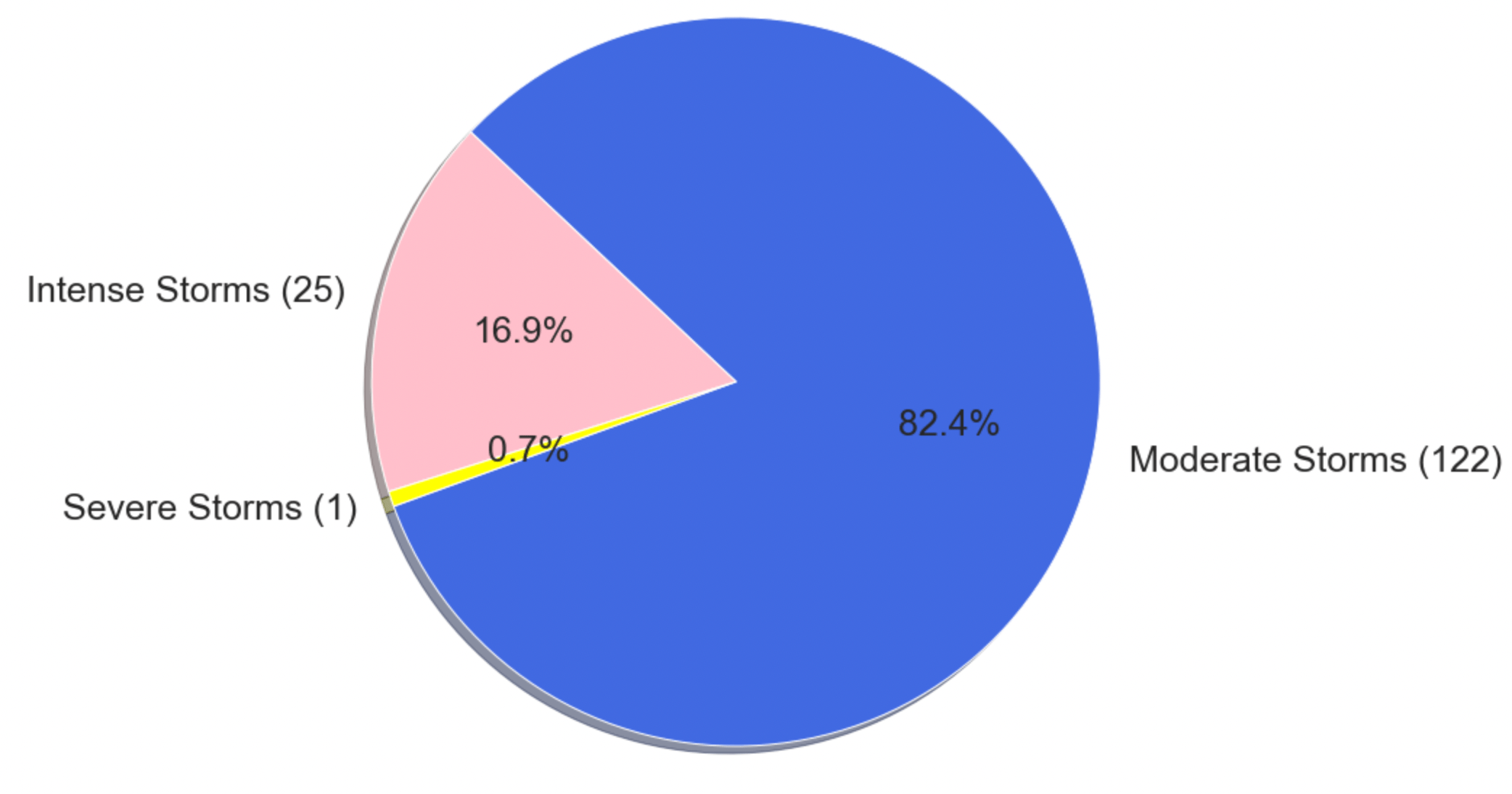} \\
   \vspace{0.2cm}
   \includegraphics[width=0.45\textwidth]{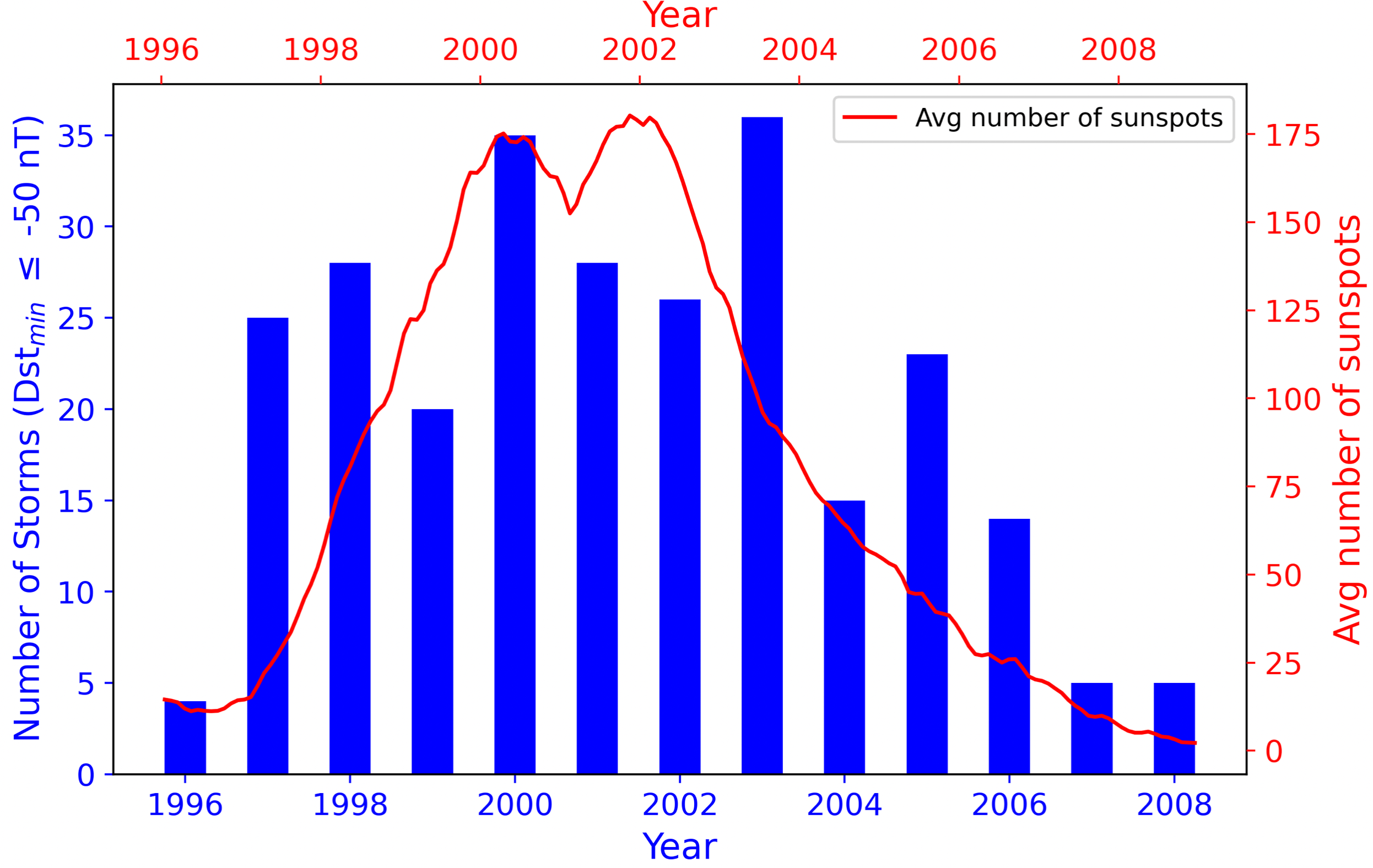}
   \hspace{0.5cm}
   \includegraphics[width=0.45\textwidth]{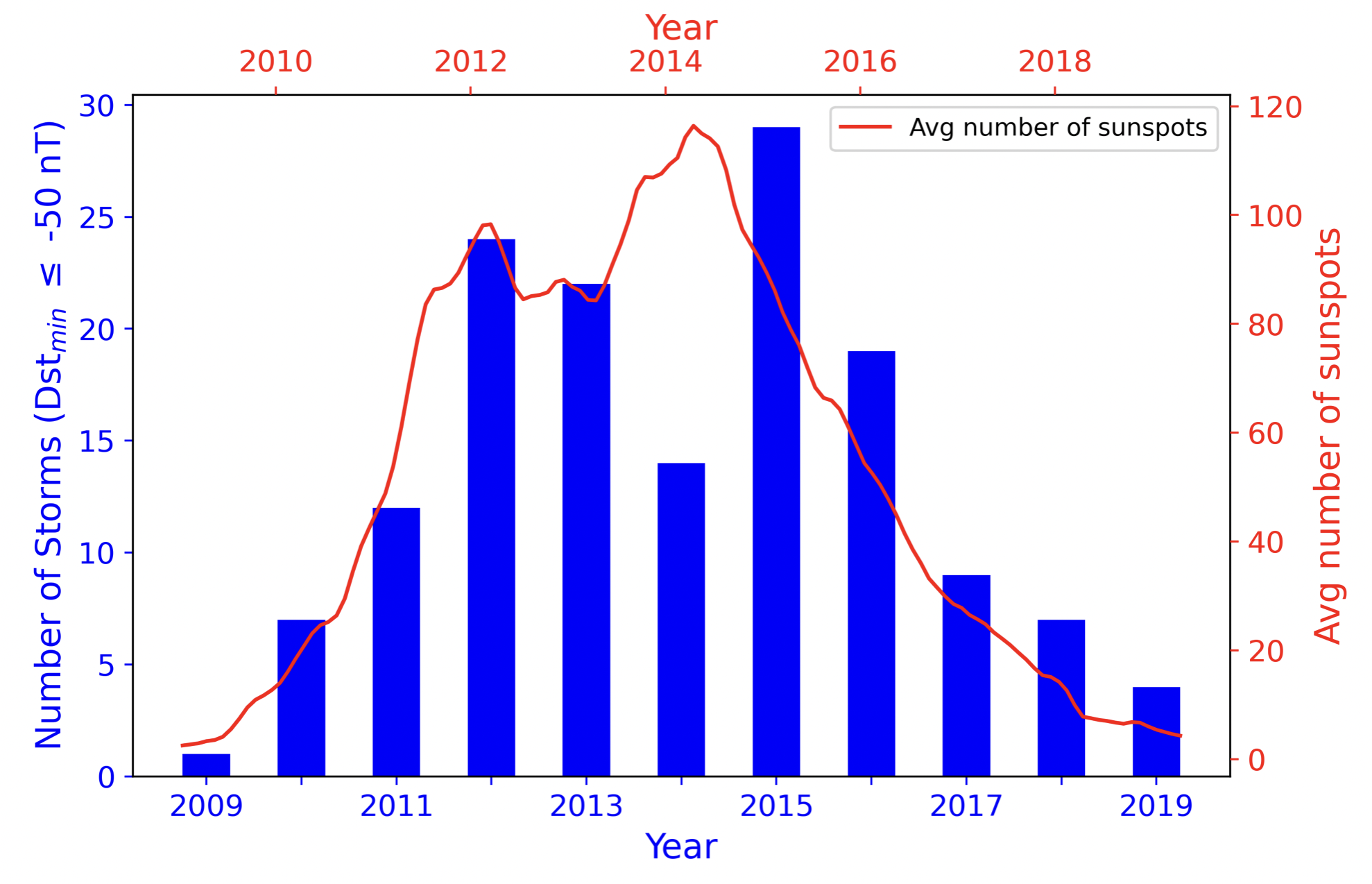} \\
   \vspace{0.2cm}
   \includegraphics[width=0.45\textwidth]{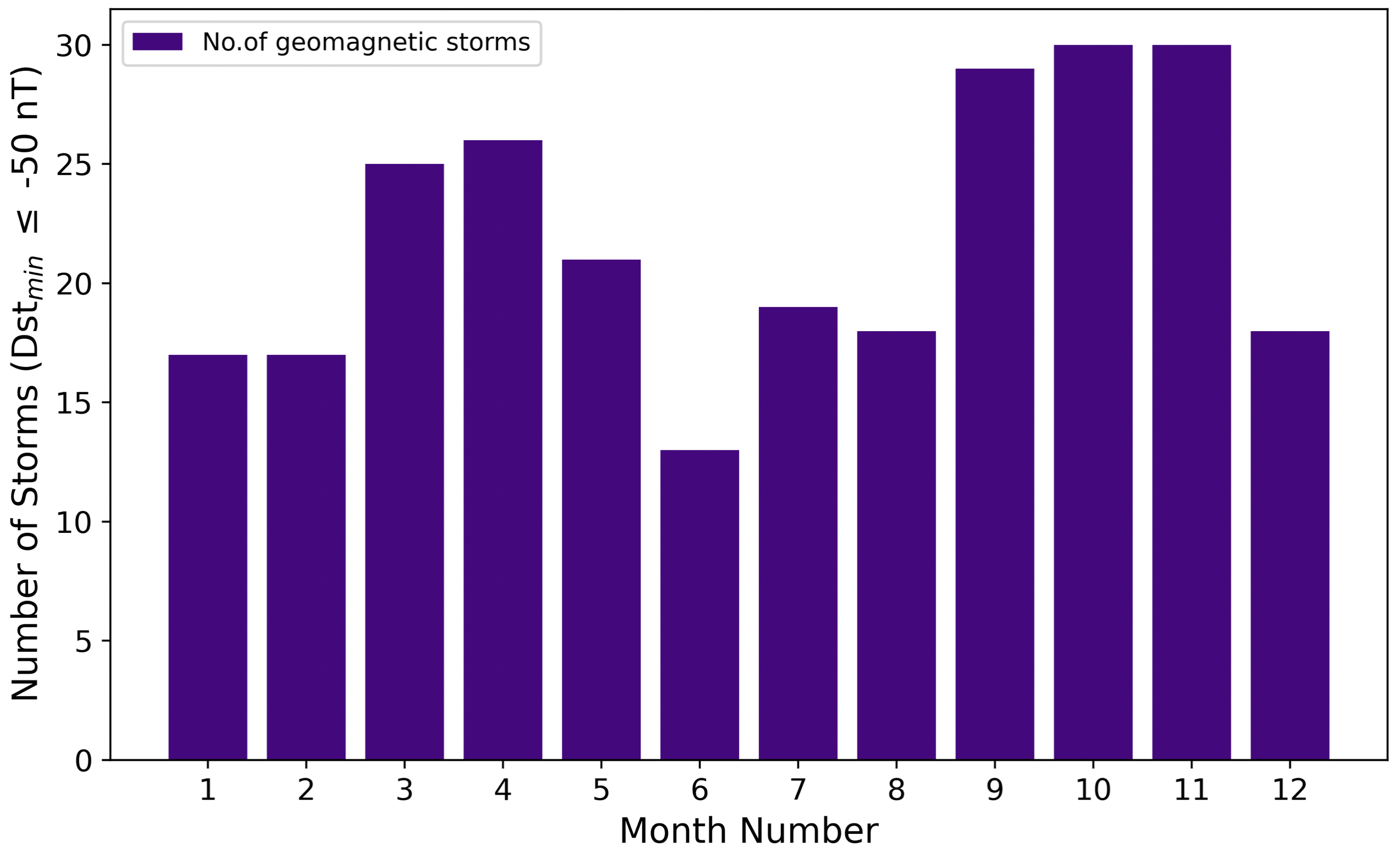}
   \hspace{0.5cm}
   \includegraphics[width=0.45\textwidth]{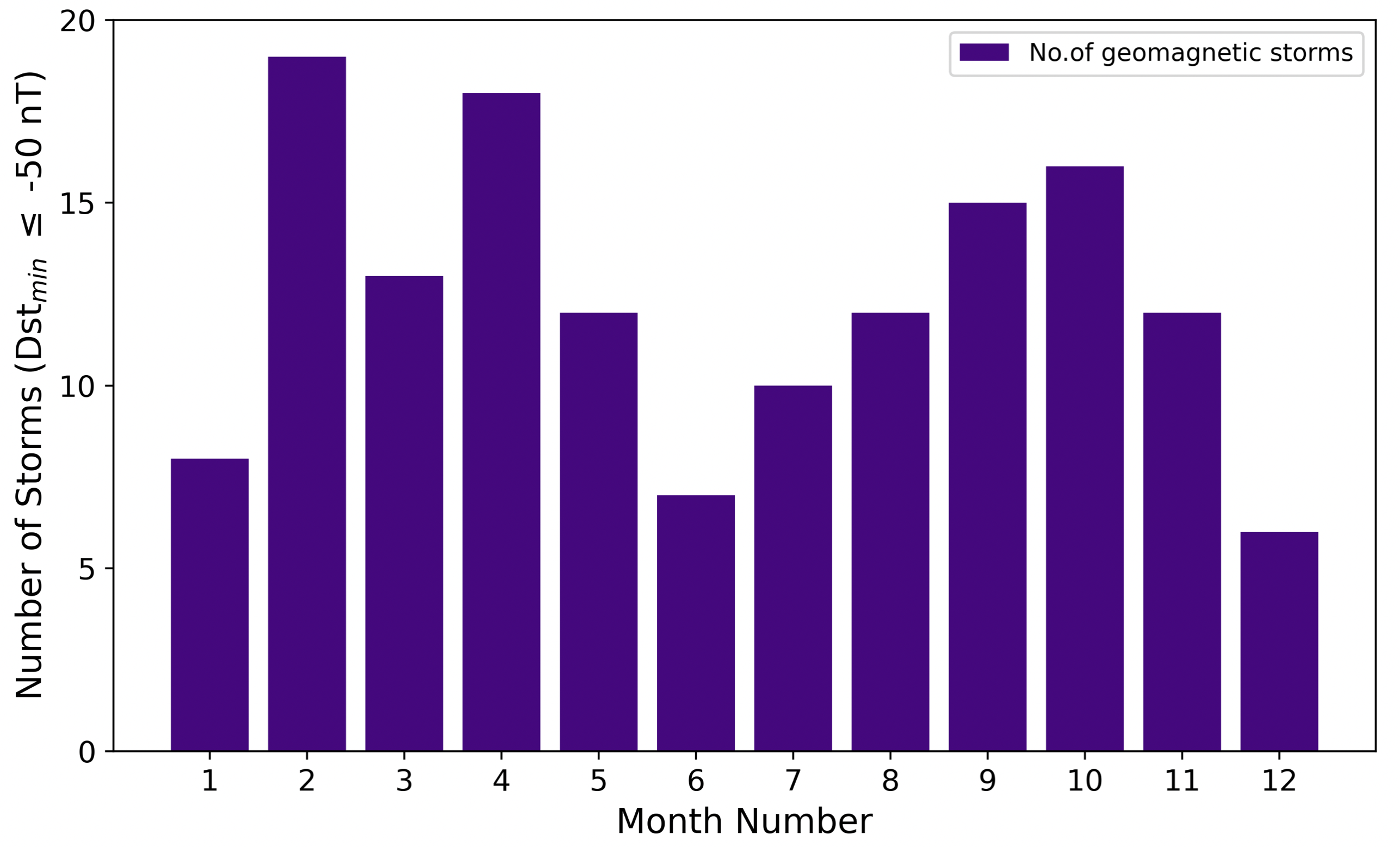}
   
   \caption{Top panel: The distribution of storms with $\text{Dst}_{\text{min}}\leq-50\;\text{nT}$ is shown for solar cycle 23 on the left, and solar cycle 24 on the right panel. The numbers in small brackets note the number of storms falling in that particular category of storms in solar cycle 23. Middle panel: The bar plots (blue) represent the number of geomagnetic storms with Dst$_{\text{min}}\leq \text{-50}\;\text{nT}$ for each year from 1996 to 2008 in solar cycle 23 in the left and from 2009 to 2019 for solar cycle 24 in the right. The line plots (red) represent 13-month smoothed total sunspot numbers. Bottom panel: Monthly distribution of the total number of storms for solar cycles 23 and 24 in the left and right panels, respectively.}
   \label{disticycle2324_50}
 
\end{figure*}

\subsection{Storms (Dst \texorpdfstring{$\boldsymbol{_{\text{min}}}$ $\boldsymbol{\leq-50}$} nT) in Solar Cycles 23 and 24 } 

The weak storms capable of almost negligible effects in geomagnetic fields are more frequent than stronger storms, and identifying their different phases and associated solar drivers is quite cumbersome. We exclude the weak storms from our statistics and focus on the distribution of all stronger storms. Figures~\ref{disticycle2324_50} show the distribution of geomagnetic storms, stronger than weak storms, in both solar cycles. As we notice from the top panel of the figure, most of the storms are moderate, while the great storms are least in number. The figure's top-left panel displays that the total number of storms in solar cycle 23 is 263, comprising 190 (72.2\%) moderate storms, 58 (22.1\%) intense storms, 11 (4.2\%) severe storms, and 4 (1.5\%) great storms. The figure's top-right panel for solar cycle 24 shows that the total number of storms is 148, comprising 122 (82.4\%) moderate storms, 25 (16.9\%) intense storms, and 1 (0.7\%) severe storms. The decrease in the total number of storms with a smaller fraction of stronger storms in solar cycle 24 than in solar cycle 23 shows a reduced solar wind-magnetosphere coupling in cycle 24. This is consistent with earlier studies, which also showed that the primary reason for reduced geomagnetic activity in cycle 24 is the lack of stronger and long-duration ICMEs and MCs with southward IMF \citep{Kilpua2014,Watari2017,Mishra2021a}.

We also examine the relationship between the 13-month smoothed monthly sunspot number and the number of storms. The 13-month smoothed monthly sunspot number is derived by a ``tapered-boxcar'' running mean of monthly sunspot numbers over 13 months centered on the corresponding month. The middle panel of Figure~\ref{disticycle2324_50} shows a distribution of the averaged monthly sunspot number and yearly number of geomagnetic storms in both solar cycles 23 and 24. For solar cycle 23 in the middle-left panel, we note two peaks in the sunspot number around 2000, coinciding with a peak in number of storms, and another more prominent peak in 2002, not coinciding with the number of storms. The number of geomagnetic storms was maximum in 2003, whereas the second highest number of storms was in 2000. In the middle-right panel of the figure, for solar cycle 24, there are two peaks in the sunspot number: one in 2012 coinciding with a peak in the number of storms, and another more significant in 2014 that does not precisely coincide with the peak in the number of storms. The double peaks in the sunspot cycle, popularly known as Gnevyshev peaks, are well-known features of the solar cycle and have been modeled in earlier studies \citep{Norton2010,Karak2018}. In cycle 24, the number of geomagnetic storms was the maximum in 2015, whereas the second highest number was in 2012.

In general, the yearly number of storms generally follows the rise and decline trends of the solar cycle strength. However, there are two clear peaks in the geomagnetic activity in both solar cycles: the first peak appears at the first peak in the sunspot cycle, while the second peak is almost a year after the second peak in the sunspot cycle. The second peak in the number of storms is in the early declining phases of the cycles, and thus, there is no perfect correlation between sunspot number and the occurrence rate of moderate and stronger than moderate geomagnetic storms. This could be possible as many moderate storms are caused by high-speed streams, SIRs, and high-latitude CMEs having their solar sources in the non-sunspot regions. It has been previously suggested that CMEs primarily contribute to the first peak in the number of storms in the solar maxima, while fast streams from coronal holes mostly cause the second peak \citep{Echer2008}. We also noted that during the overlap (deep minimum between 2007-2009) of cycles 23 and 24, the geomagnetic activity was the lowest, consistent with the lowest number of sunspots observed then. The geomagnetic storms are fewer in cycle 24 than in the respective phases of cycle 23. This is expected as the sunspot numbers and total mass loss rate via CMEs at the peak of cycle 24 are reduced by around 40\% and 15\%, respectively, to that at the previous cycle's peak \citep{Mishra2019}.

The rate of CMEs and CIRs dependent on the 11-year solar activity cycle can modulate the occurrence rate of geomagnetic storms. In addition, it is also possible that the Earth's orbit around the Sun can broadly impact the geomagnetic activity variation. Earlier studies have shown more frequent occurrences and higher strength of geomagnetic activity during equinoxes, which are explained using three mechanisms: axial hypothesis, equinoctial hypothesis, and Russell‐McPherron effect \citep{Cliver2000,OBrien2002,Lockwood2020}. We examined the monthly distribution of all the storms with $\text{Dst}_{\text{min}}\leq-50\;\text{nT}$ to understand how they are distributed month-wise for solar cycles 23 and 24. From the bottom-left panel of Figure~\ref{disticycle2324_50}, it is clear that the number of geomagnetic storms is highest from September to November and then second-highest from March to April in solar cycle 23. The bottom-right panel shows that in solar cycle 24, there are many storms from February to April and then from September to October. We note that geomagnetic storms at equinoxes can reach up to twice the number of storms at solstices around January and June. The most prominent peak in the monthly distribution of the number of storms in solar cycle 23 is soon after the September equinox while slightly before the March equinox in solar cycle 24. However, \citet{Mursula2011} has reported that semiannual variation in global geomagnetic activity is overestimated, and it is an artifact of the dominant annual variation with maxima alternating between Spring and Fall in consecutive cycles.

\begin{figure*}

 \centering
   \includegraphics[width=0.45\textwidth]{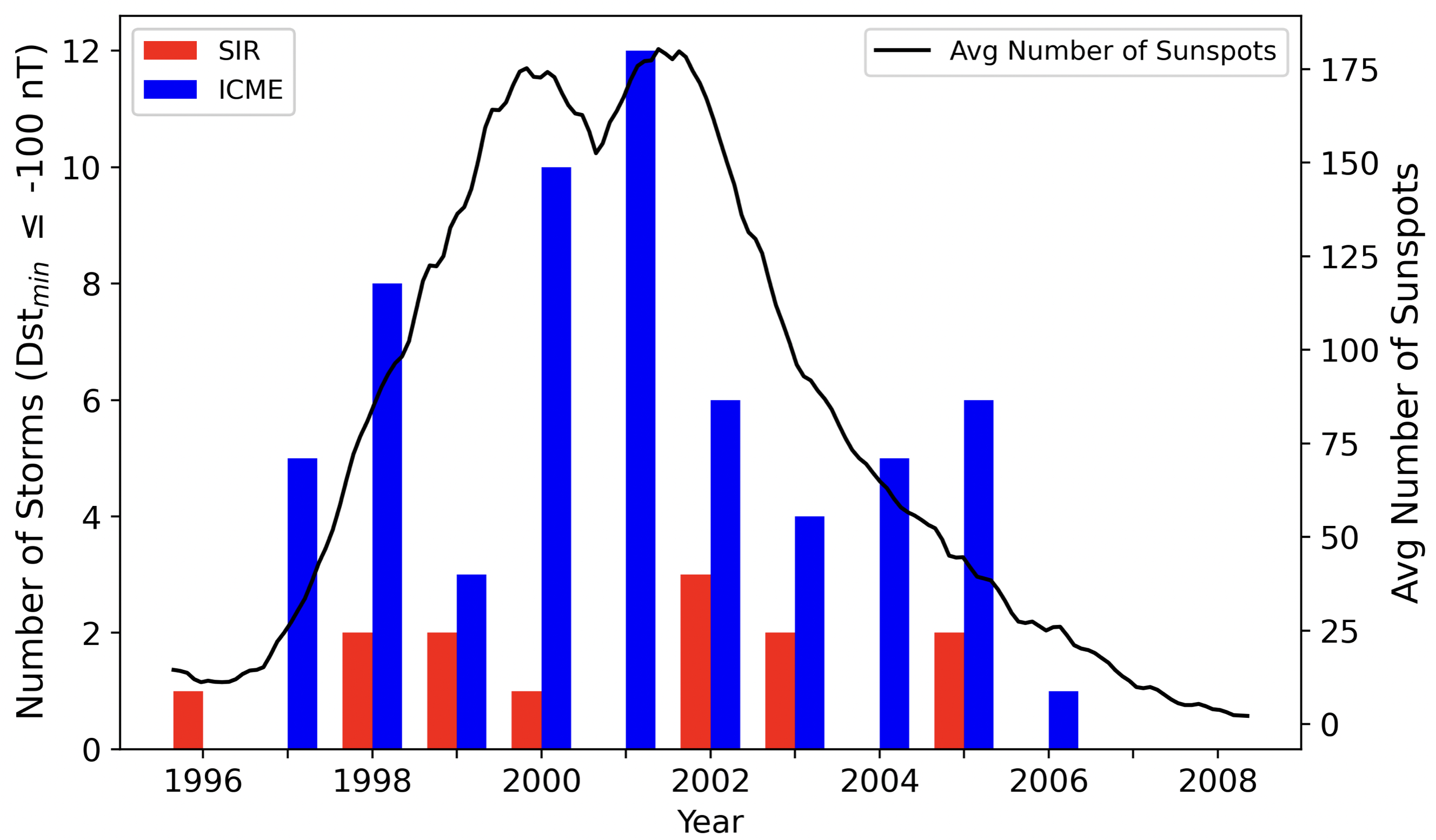}  
   \hspace{0.5cm}
   \includegraphics[width=0.45\textwidth]{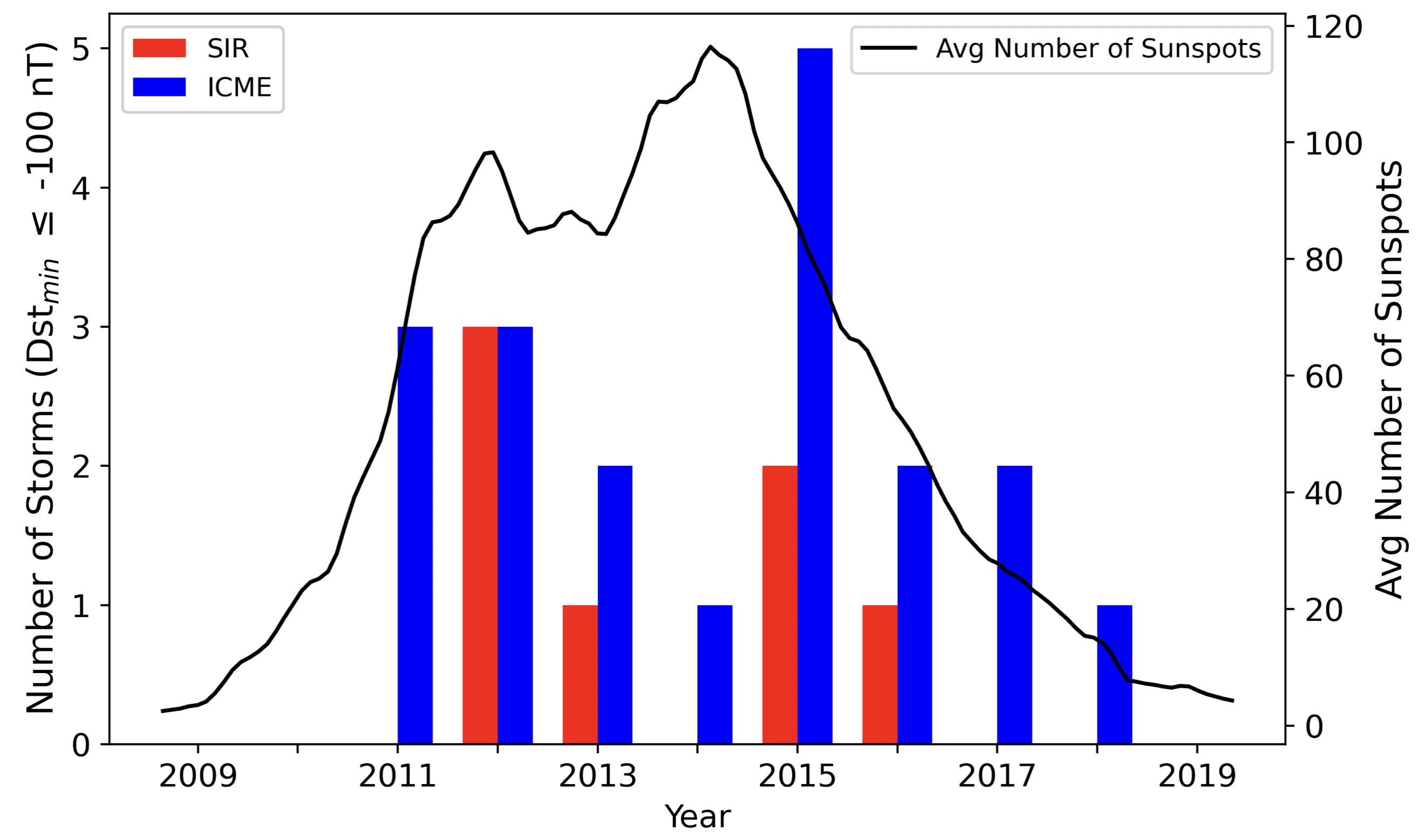} \\
   \vspace{0.2cm}
   \includegraphics[width=0.45\textwidth]{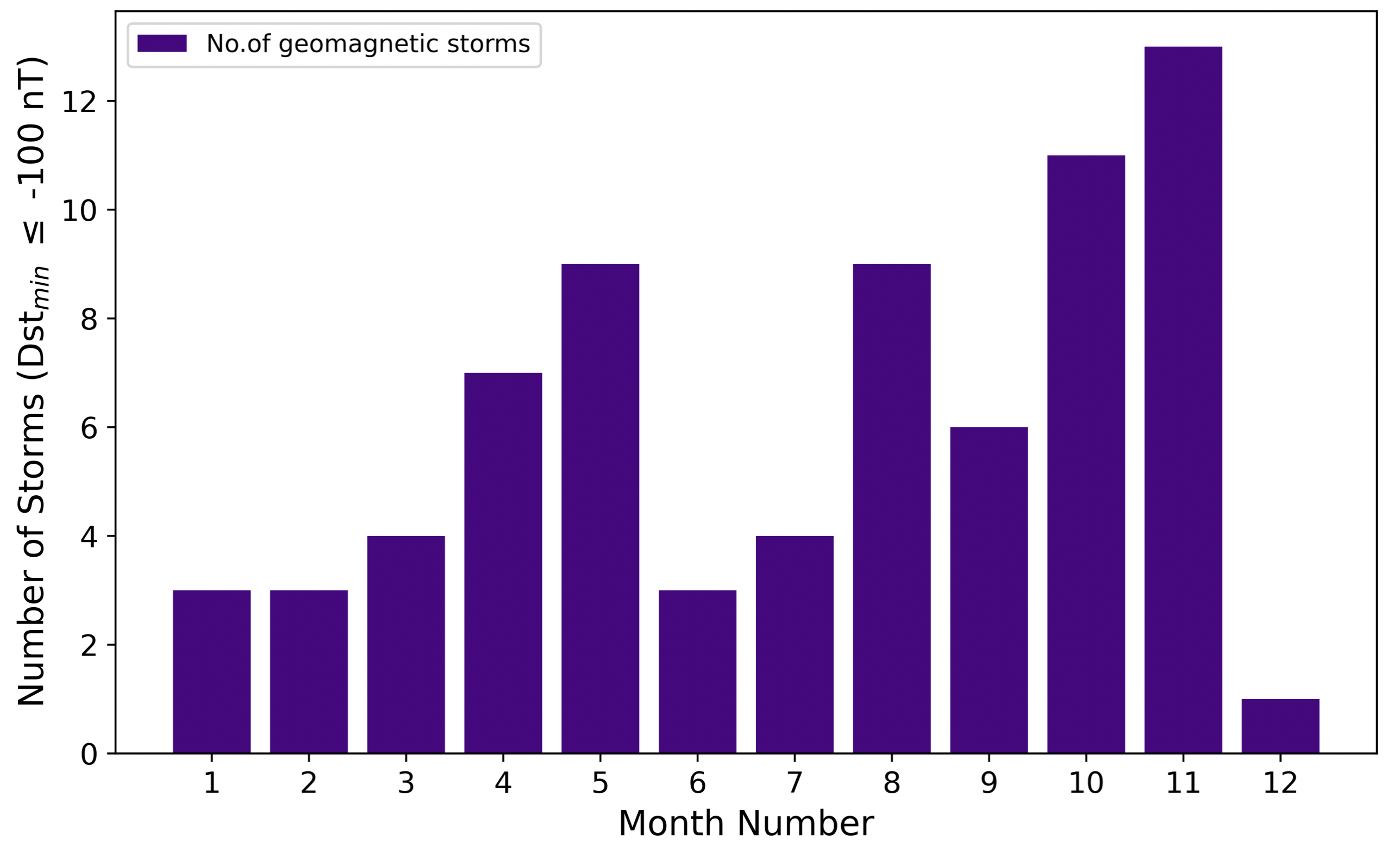}
   \hspace{0.5cm}
   \includegraphics[width=0.45\textwidth]{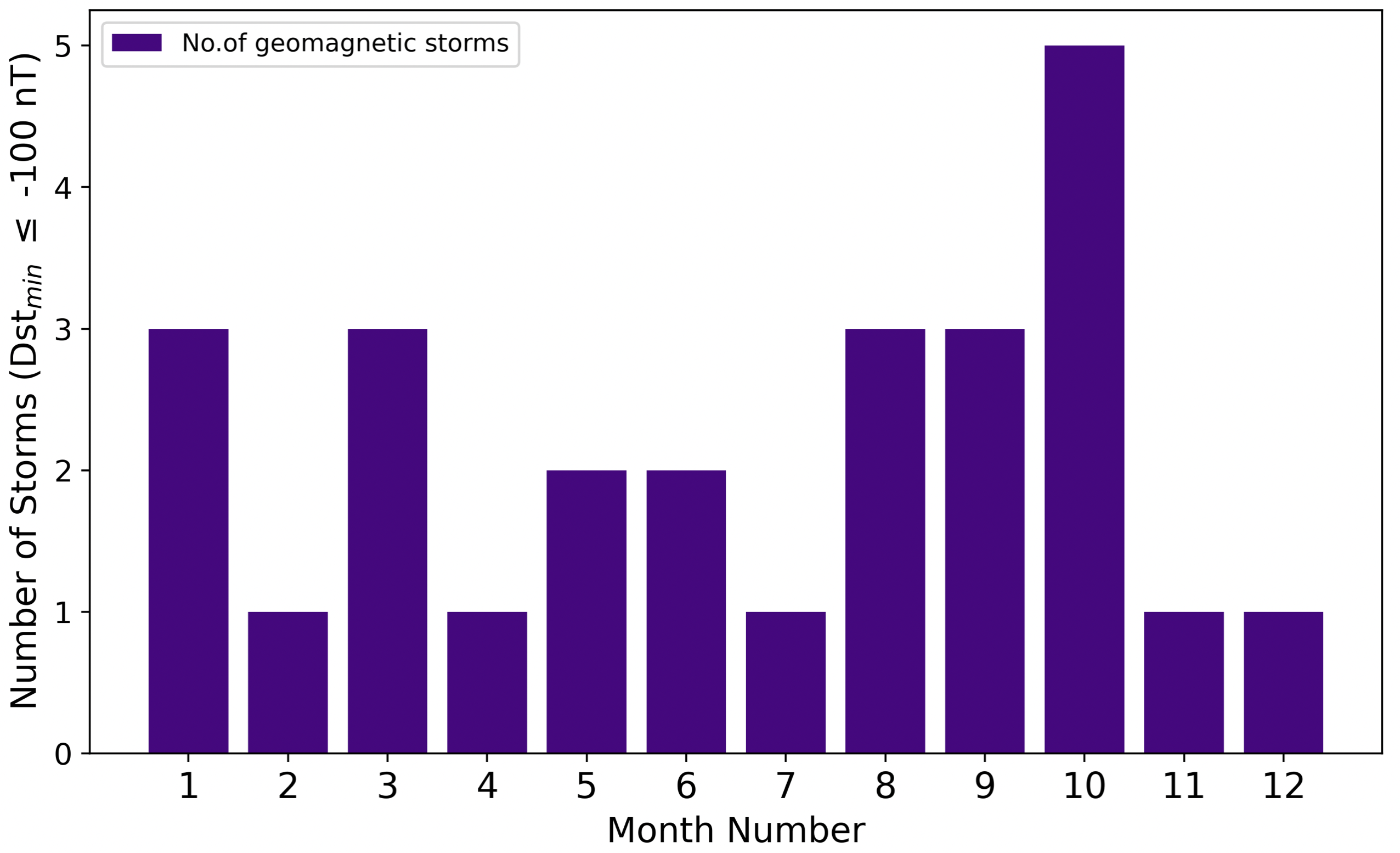}
   
   \caption{Top panel: The bar plots show number of geomagnetic storms with Dst$_{\text{min}}\leq \text{-100}\;\text{nT}$. for each year from 1996 to 2008 in solar cycle 23 in the left and from 2009 to 2019 for solar cycle 24 in the right. The blue and red bars represent the storm's drivers as ICME and SIR, respectively. Bottom panel: Monthly distribution of the total number of storms for solar cycles 23 and 24 in the left and right panels, respectively.}
   \label{disticycle2324_100}
 
\end{figure*}

\subsection{Storms (Dst\texorpdfstring{$\boldsymbol{_{\text{min}}}$ $\boldsymbol{\leq-100}$} nT) in Solar Cycles 23 and 24 } 
\label{dstlt100_2324}

This section delineates the primary causative factors behind intense and stronger-than-intense geomagnetic storms, specifically attributing them to coronal mass ejections (CMEs), stream interaction regions (SIRs), or composite structures resulting from their mutual interactions. Therefore, we exclude events categorized as weak or moderate storms, as their relatively larger frequency can potentially overshadow our detailed scrutiny of intense storms. Notably, within solar cycle 23, there are 73 storms with a $Dst\leq-100$ nT, while solar cycle 24 records a considerably lower count of 26 such storms. This discrepancy underscores the diminished geomagnetic activity during solar cycle 24. Our investigation encompasses an examination of the fluctuations in sunspot numbers vis-à-vis the annual frequency of intense or stronger-than-intense geomagnetic storms.

We identify interplanetary drivers for each storm using three catalogs: Richardson and Cane catalog for ICMEs \citep{Cane2003,Richardson2010}, ICME catalog by \citet{Shen2017} and SIR catalog by \citet{Chi2018}. We analyzed the sources of each storm and categorized them, irrespective of single-peaked or multiple-peaked in Dst (discussed in the next section), into four distinct classes: (i) storms driven by isolated ICME referred as I-ICME, (ii) those driven by multiple interacting ICMEs referred as M-ICME, (iii) driven by SIR, and (iv) driven by interactions between ICMEs and SIRs referred as ICME-SIR. In solar cycle 23, we found that out of 73 storms with a $Dst\leq-100$ nT,  37 storms were produced by I-ICMEs, 23 by M-ICMEs, 9 by SIR, and 4 by the interacting ICME-SIR drivers. In solar cycle 24, out of 26 storms with a $Dst\leq-100$ nT, there are 15 storms due to I-ICME, 4 storms by M-ICME, 2 by SIR, and 5 storms resulting from the interaction between ICME and SIR. To confer statistical robustness upon our analysis, we further streamlined the four distinct categories of storm drivers into a binary classification: (i) storms driven by ICMEs: either singular or multiple ICMEs, and (ii) storms driven by SIRs: either from SIRs or ICME-SIR interacting structure. Although SIRs alone can drive intense geomagnetic storms, we acknowledge that ICME could be a main driver for some storms classified under the interacting ICME-SIR category. There are studies about intense geomagnetic storms primarily driven by ICME compressed in a CIR ahead of a high-speed solar wind stream \citep{Nitta2017,Nitta2021}. However, our current study considered categorizing storms driven by ICME-SIR into the SIR category and did not specifically analyze the main driver between SIR and ICME. Since there are only around 5\% and 20\% of the total number of storms attributed to the ICME-SIR category in solar cycles 23 and 24, respectively, our classification approach does not alter the overall statistical findings derived from the binary classification of storm drivers.

Figure~\ref{disticycle2324_100} graphically represents the distribution of ICME-driven and SIR-driven annual storm occurrences alongside sunspot numbers. Remarkably, in solar cycle 23, the peak of ICME-driven geomagnetic storms is observed during 2000-2001, closely aligned with the dual peaks in sunspot activity in 2000 and 2002. However, the SIR-driven number of storms is almost the same throughout cycle 23, contributing to around 20\% of the total intense storms in the rise and decline phase, which reduced to less than 10\% during the cycle's maximum. Examining the right panel of Figure~\ref{disticycle2324_100}, we note that the maxima of ICME-driven geomagnetic storms within solar cycle 24 occurred in 2015, a year after the second peak in the sunspot numbers. A noteworthy observation is that SIRs-driven intense storms in solar cycle 24 comprise up to 30\% of the total number of storms in the maximum of the cycle, and it could comprise as much as ICMEs in the rising phase of the cycle. SIR-associated storms are indeed larger in the weaker solar cycle 24 than in the cycle 23.

The bottom panel of Figure~\ref{disticycle2324_100} shows the distribution of intense and stronger-than-intense geomagnetic storms over months. For cycle 23, shown in the left-bottom panel, the semiannual variation with maximum intensity is noted around May and November, i.e., two months after the equinoxes. However, for solar cycle 24, the number of storms peaks around March and October. The most prominent peak in the distribution of storms is soon after the September equinox for both cycles 23 and 24.


\section{Analysis of the Recovery Phase of Geomagnetic Storms (Dst $_{\text{min}}\leq $ 100)} 

\begin{figure*}
    
        \centering
        \includegraphics[width=0.47\textwidth]{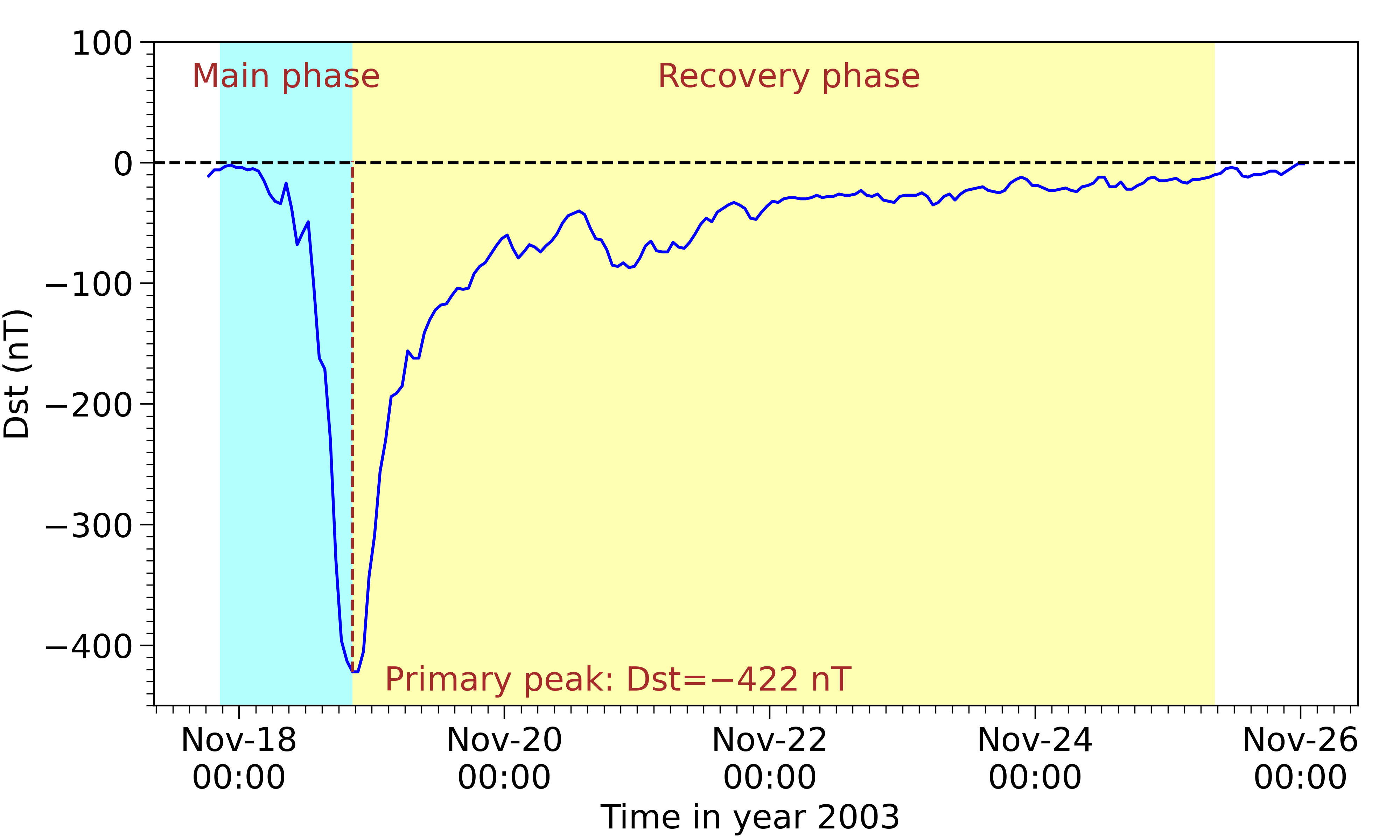} 
        \hspace{0.3cm}
        \includegraphics[width=0.47\textwidth]{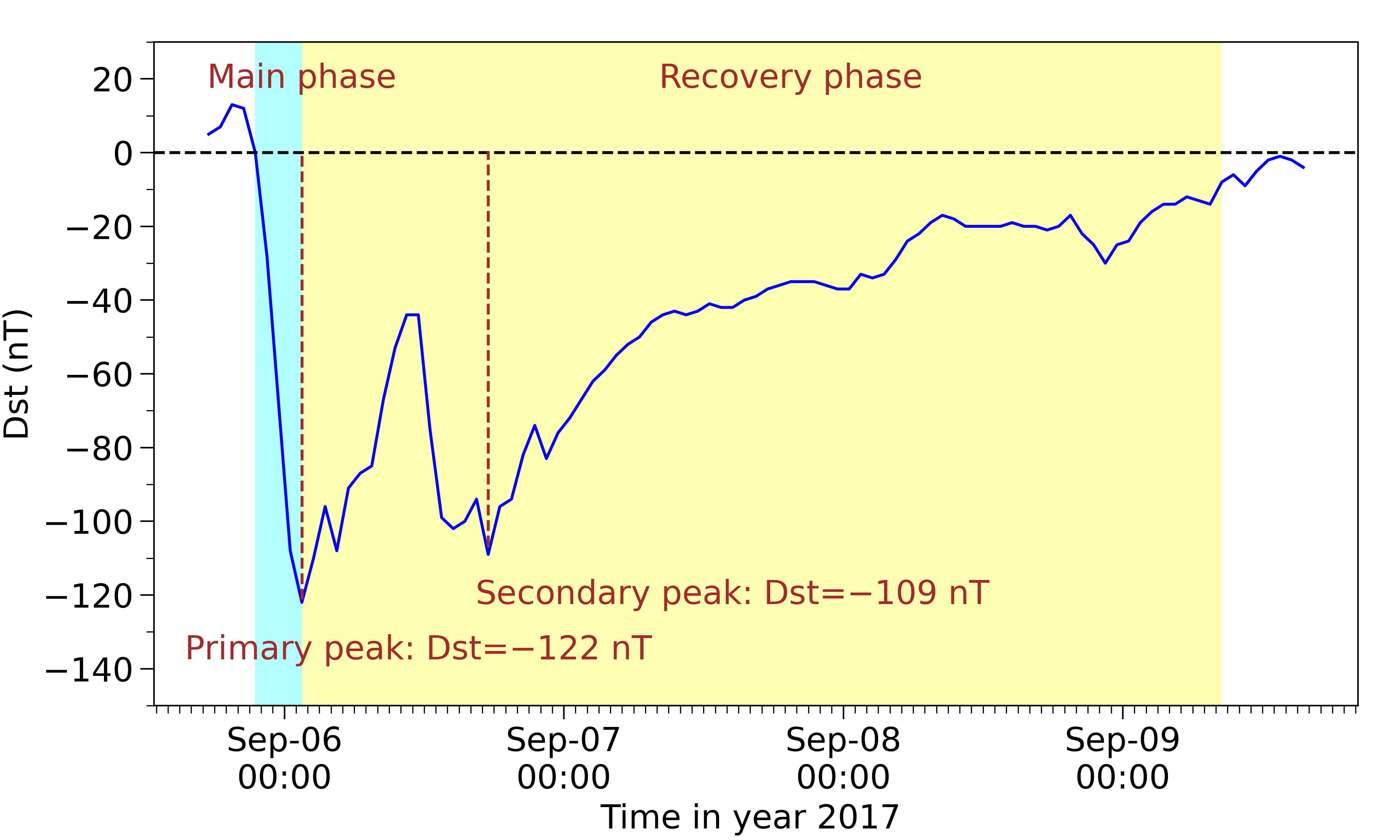} \\
        \vspace{0.5cm}
        \includegraphics[width=0.4\textwidth, trim={0.2cm 1.7cm 0.2cm 0.2cm},clip]{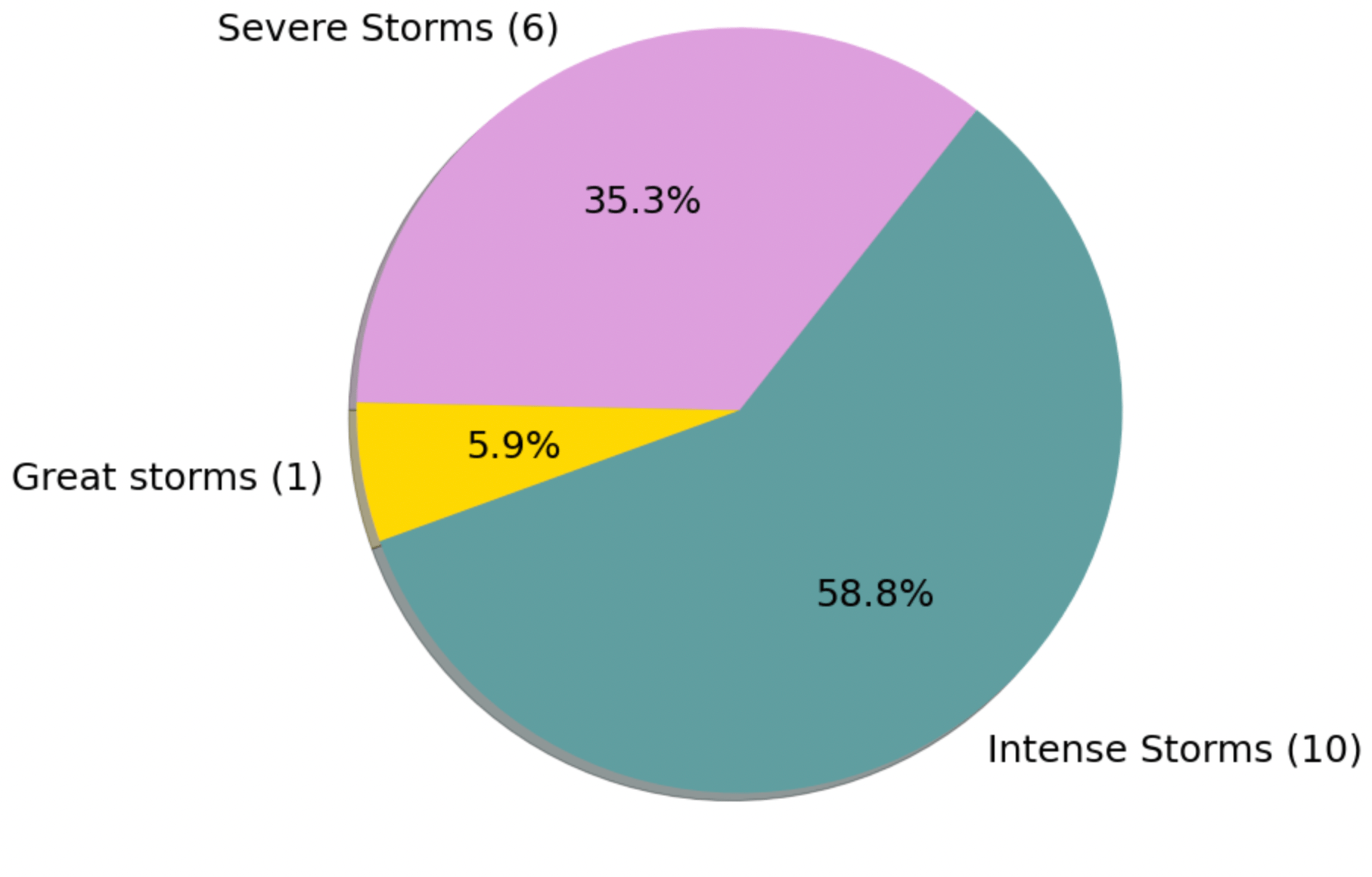}
        \hspace{0.9cm}
        \includegraphics[width=0.45\textwidth, trim={0.2cm 0cm 0.2cm 0.2cm},clip]{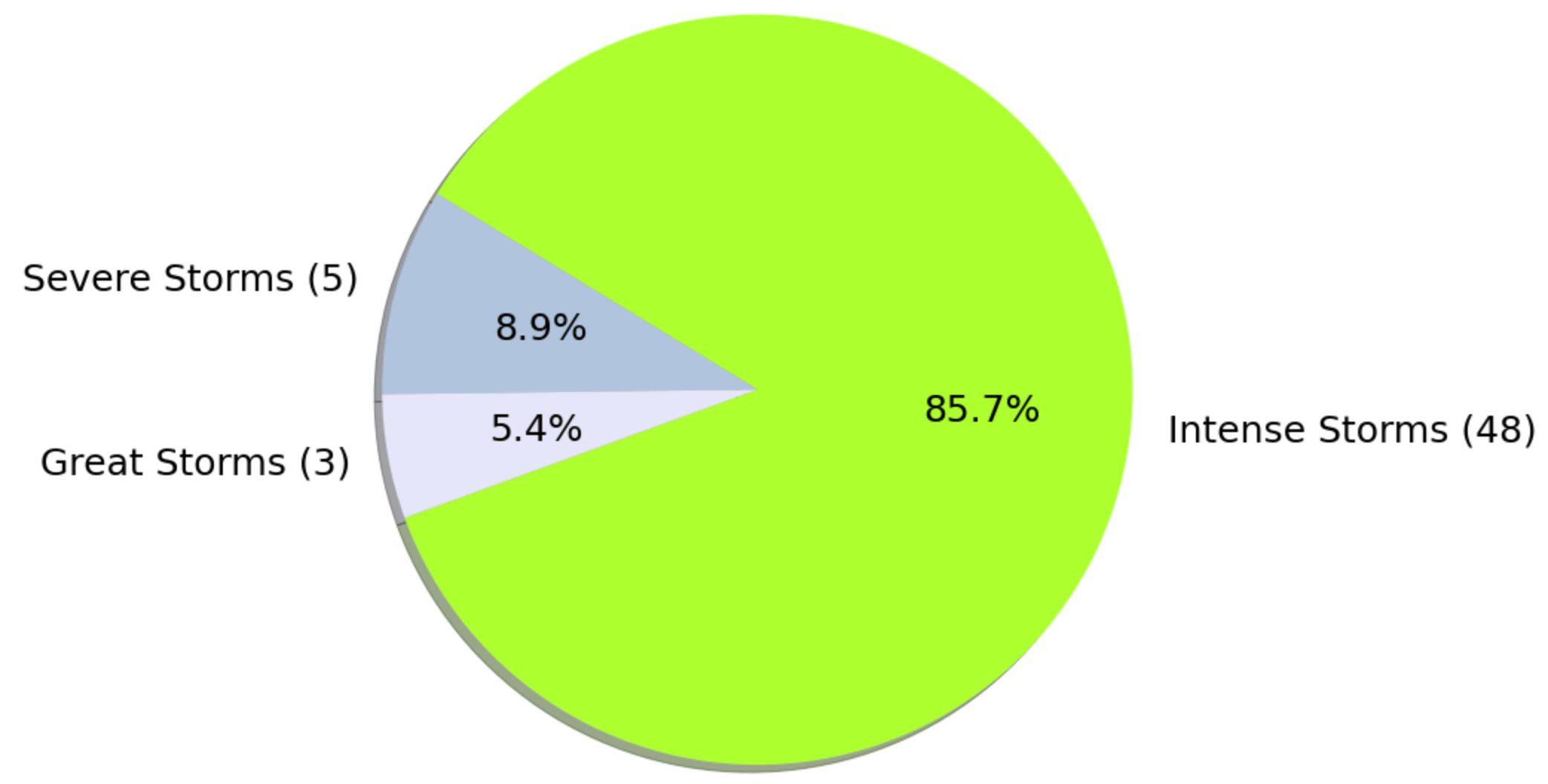}  \\
        \vspace{0.5cm}
        \includegraphics[width=0.45\textwidth]{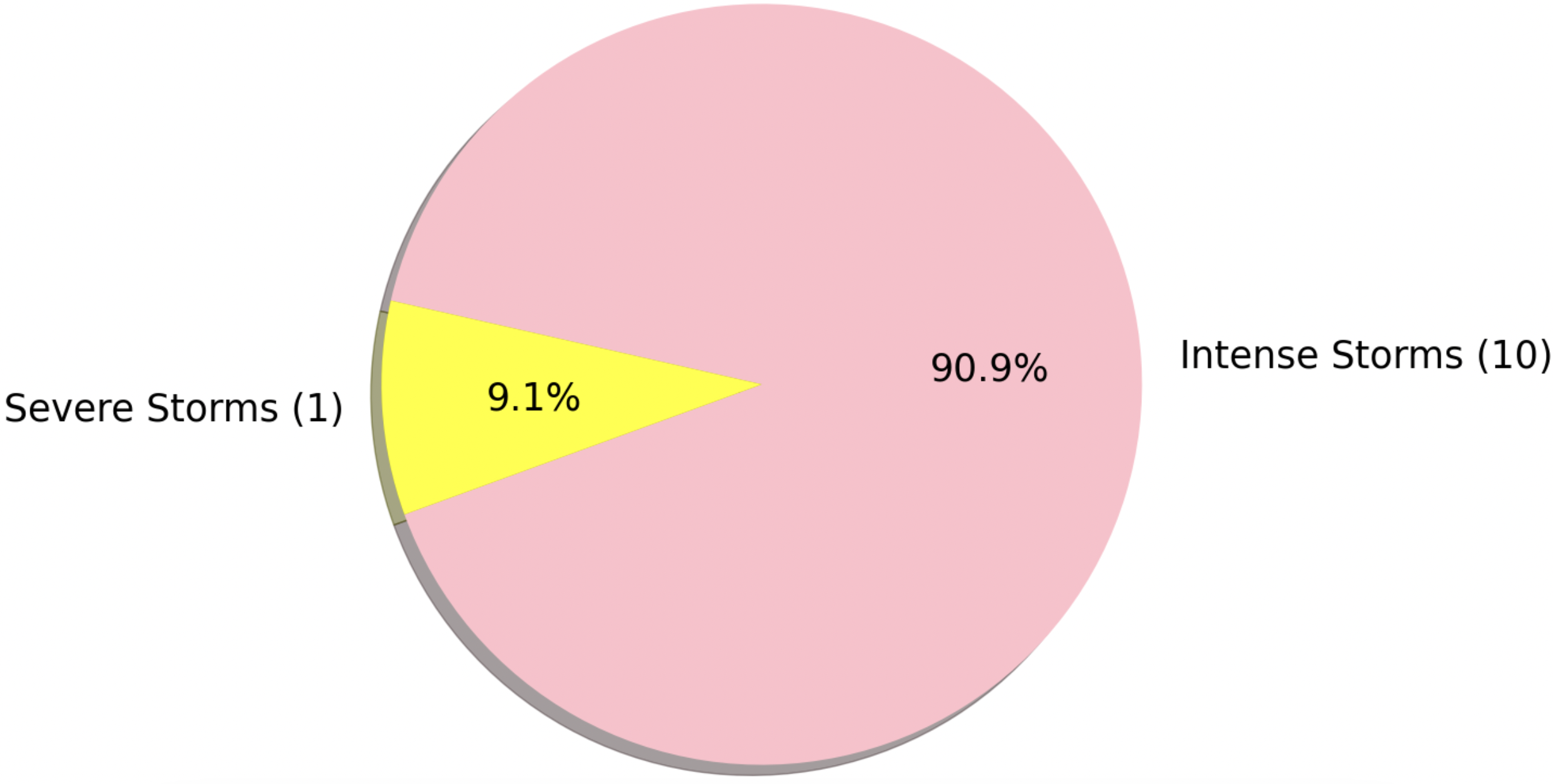}
        \includegraphics[width=0.45\textwidth]{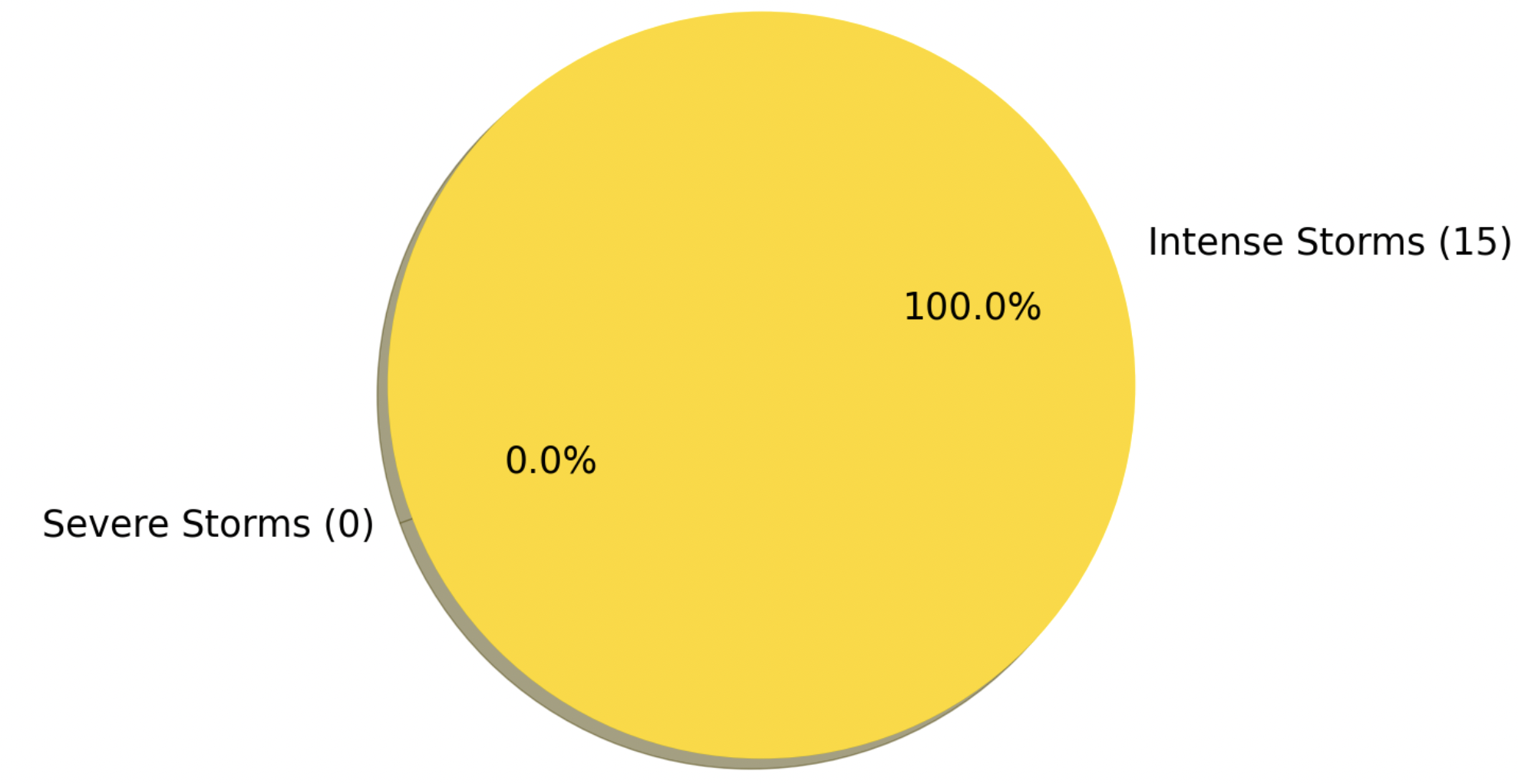}
       \caption{Top panel: An ideal single-peaked great geomagnetic storm occurred in November 2003 with its magnitude of $-422\;\text{nT}$, having smooth main and recovery phases is shown on the left. Multiple-peaked storm in September 2017, with the primary Dst peak reaching $-122\;\text{nT}$ is shown on the right. The shaded areas with cyan and yellow show the duration of the main and recovery phases for both the single and multiple-peaked storms. Middle panel: This pie chart represents the distribution of single-peaked and multiple-peaked storms during solar cycle 23 in the left and right panels. Bottom panel: It is the same as the middle panel, but for solar cycle 24.}
        \label{sm_storms2324}
   
\end{figure*}

Analyzing the recovery phase of geomagnetic storms and comparing it with the storm's characteristics is instrumental in comprehending the decay of the magnetospheric ring current's intensity. It is worth highlighting the challenge of precisely delineating the start and end points of a geomagnetic storm's recovery interval, given the potential existence of multiple peaks during the main and recovery phases. An illustrative instance of single and multiple-peaked storms is provided in the top panel of Figure~\ref{sm_storms2324}. The top-left panel of the figure showcases an isolated storm where only a single Dst peak is observed. However, in scenarios involving multiple-peaked storms, as shown in the top-right panel of the figure, two (primary and secondary) Dst peaks can manifest. Our developed automated algorithms allow us to distinguish between isolated and multiple-peaked storms.

\subsection{Classification of single-peaked and multiple-peaked Storms}

The categorization of storms into single-peaked and multiple-peaked storms depends on the chosen threshold for identifying secondary peaks during the storm's main or recovery phase. To identify storms with multiple peaks, \citet{Kamide1998} used the criteria that the intervening recovery should not exceed 90\% of the first peak and that the peaks should be separated by more than 3 hours. In our study, we first find the Dst minimum values associated with the storms and segment the storm interval into its main and recovery phases. Our automated algorithms facilitate the identification of Dst minimum values for storms falling into the category of intense or stronger-than-intense (where Dst$_{\text{min}}\leq \text{-100}\; \text{nT}$). We utilize the lowest (most negative) Dst value among all peaks for multiple-peaked storms as the pivotal criterion for classifying storms into distinct intensity levels.

Subsequently, we extract sub-intervals from both the main and recovery phases, encompassing values ranging between 40\% and 95\% of the most significant (i.e., primary) Dst peak index associated with the identified storm. We look for additional peaks within the extracted sub-intervals characterized by substantial and sustained fluctuations. The criteria for identifying such substantial and sustained peaks alongside the primary Dst peak of the storm are as follows: a fluctuation (reversing trend) equal to or less than -5 nT within an hour, followed by the same consistent trend of increase or decrease for at least five hours. However, we recognize that our scheme will not identify some storms with closer-spaced peaks. Still, we think such inaccuracies will be less likely for intense storms, which is the focus of our present study. The rationale for classifying a storm as multiple-peak is rooted in the unlikelihood of Dst reversing its trend without a significant change in the geoeffective parameters associated with different sub-structures of a single ICME/SIR or multiple interplanetary drivers \citep{Farrugia2006}. By applying these criteria, we successfully differentiate between single-peak and multiple-peak storms during solar cycles 23 and 24.

\subsection{Distribution of single-peaked and multiple-peaked storms}

We further examine the distribution of intense and stronger-than-intense storms into distinct categories based on their intensity relative to the primary Dst peak values. For solar cycle 23, the distribution of single-peaked and multiple-peaked storms is depicted on the left and right in the middle panel of Figure~\ref{sm_storms2324}. The equivalent distribution for solar cycle 24 is illustrated in the bottom panel of the same figure, again with left and right panels representing single-peaked and multiple-peaked storms. Evidently, within solar cycle 23, most (10) single-peaked storms fall into the intense category, while fewer (6) are classified as severe, and the least (1) belong to the great storms category. In contrast, solar cycle 24 exhibits only 11 intense storms, one severe storm, and no great storms, indicating a weaker solar cycle than its predecessor. For multiple-peak storms, cycle 23 displays 48 intense storms, 5 severe storms, and 3 great storms. However, in cycle 24, there are only 12 intense storms with no severe or great storms. This pattern underscores the relationship between a stronger sunspot cycle and more potent storms, especially those with multiple peaks, possibly resulting from multiple large-scale solar wind structures depositing their energy into Earth's magnetosphere \citep{Liu2014}.

\subsection{Main phase and recovery phase parameters of storms}

We focus on examining the buildup (main phase) of storms occurring during the last two solar cycles, as well as their subsequent recovery phases. For each storm, we demarcate the end of the recovery phase as the moment when the recovered Dst (Disturbance storm time) parameter approaches values greater or equal to -10 nT after reaching the Dst peak. We calculate the time difference between the Dst peak and the recovery phase's termination to derive the recovery phase's duration. To estimate the duration of the main phase, we identify two critical points by an automated algorithm: the first point where the transition from positive to negative Dst values occurs before the Dst peak is the initiation of the main phase, and the second point at which the Dst reaches its minimum (peak) value is the end of the main phase of the storm. It's worth noting that numerous storms exhibit a main phase characterized by entirely negative Dst values. In such instances, although the algorithm could find the end of the main phase, we manually look at Dst profiles and associated storm drivers for fixing the initiation time of the main phase. The top-left panel of Figure~\ref{sm_storms2324} shows a typical example of the main phase and recovery phase duration with cyan and yellow, respectively, of a single-peaked storm. The top-right panel shows the duration of the main phase (cyan) and recovery phase (yellow) for a typical multiple-peaked storm. Concerning the recovery phase duration, we subdivide it into two segments known as the fast decay phase and the slow decay phase \citet{Aguado2010,Cid2013,Yermolaev2016}. The fast decay phase corresponds to the interval during which a storm recovers by 75\% of the minimum value of the Dst index. Following the defined fast decay phase, a storm gradually returns to ambient conditions in a slow decay phase. The duration of the slow decay phase is computed as the difference between the overall recovery phase duration and the fast decay period.

To understand the storm main and recovery phases, we look for correlations between various storm characteristics such as main phase duration, main phase buildup rate, recovery phase duration, recovery rate, fast decay phase, etc. For example, the main phase buildup rate (nT/hrs) is determined by dividing the minimum Disturbance storm time (Dst) value by the duration of the main phase. We employ Pearson and Spearman correlation coefficients to evaluate relationships between various storm characteristics during their main and recovery phases. Pearson's correlation coefficient (denoted as Pearson's $r$) quantifies the linear relationship between two datasets, while Spearman's correlation coefficient (Spearman's $r$) assesses rank correlations, facilitating comparisons between datasets with monotonically related values, even when their relationship is non-linear.

\subsection{Recovery phase of single-peaked and multiple-peaked storms of solar cycle 23}

We examine the duration of the recovery phase with other characteristics of the storms and also understand the effect of different drivers on the storms. We categorize storms, whether single-peaked or multiple-peaked in Dst, into four distinct classes based on their identified driving interplanetary structures: (i) single ICME marked as I-ICME, (ii) interacting ICMEs marked as M-ICME, (iii) SIR, and (iv) interactions between ICMEs and SIRs marked as ICME-SIR. The identification of interplanetary drivers for each storm is done using three catalogs: Richardson and Cane catalog for ICMEs \citep{Cane2003, Richardson2010}, ICME catalog by \cite{Shen2017} and SIR catalog by \cite{Chi2018}. In our sample, we note that some single-peaked storms are also associated with multiple drivers and vice-versa. Depending on the accuracy of identifying the storms into single-peaked and multiple-peaked categories, this may have some uncertainties.

\begin{figure*}

        \centering
        \includegraphics[width=0.42\textwidth]{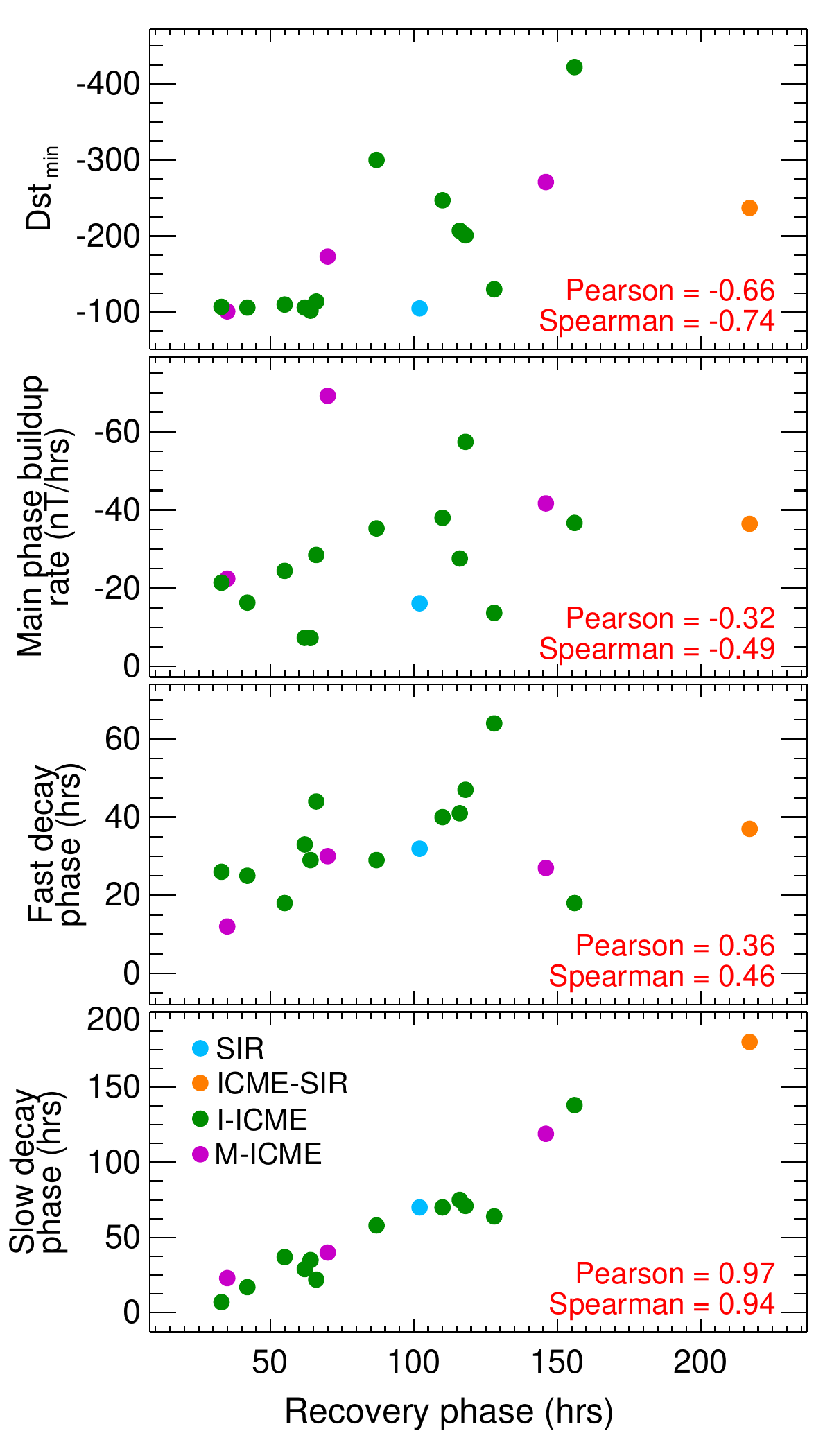}
        \hspace{8mm}
        \includegraphics[width=0.455\textwidth]{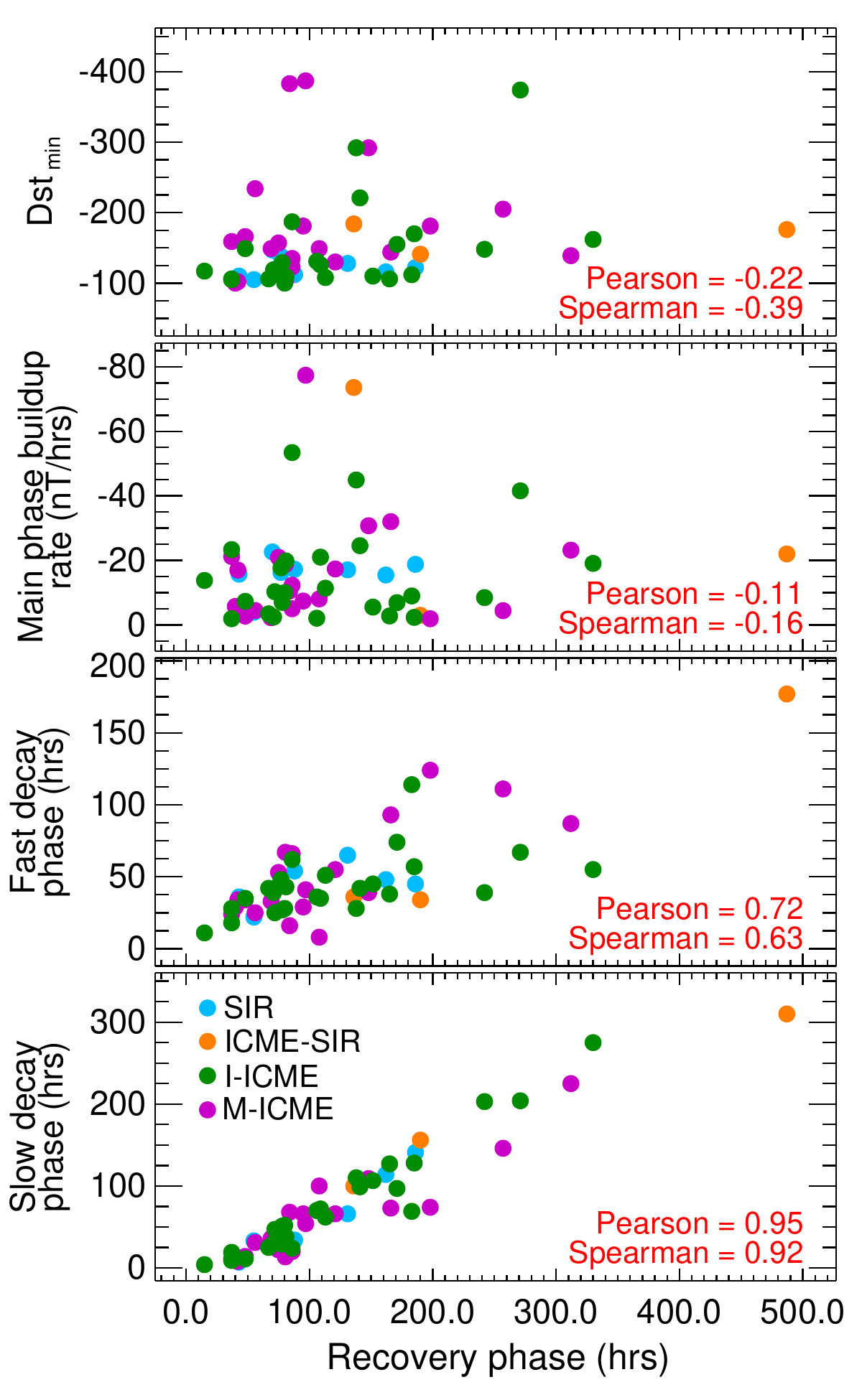}
        
        \caption{Left panel: The correlation between the recovery phase duration of single peak storms of solar cycle 23 and storm characteristics is shown. From the top to the panel, the correlation of the recovery phase with the peak Dst index, main phase buildup rate, duration of the fast decay phase, and slow decay phase, are shown, respectively. Right panel: It is the same as the left panel but for multiple peak storms.}
        \label{sc23_recov}
  
\end{figure*}

In the solar cycle 23, we note that there were 17 geomagnetic storms characterized by single peaks, with Dst peak values $\leq -100\;\text{nT}$. Out of these, 12 storms were produced by I-ICMEs, 3 by M-ICMEs, 1 by a SIR, and 1 by the interacting  ICME-SIR drivers. In contrast, there were a total of 56 storms exhibiting multiple peaks in cycle 23, with 25 attributed to I-ICMEs, 20 to M-ICMEs, 8 to SIRs, and 3 resulting from the interaction between ICMEs and SIRs. In solar cycle 24, it is noteworthy there are 11 single-peak geomagnetic storms with Dst peak values $\leq -100;\text{nT}$. Among this set, there are 6 storms due to I-ICME, 1 storm by M-ICME, 1 by SIR, and 3 storms resulting from the interaction between ICME and SIR. Furthermore, in cycle 24, 15 storms have multiple peaks, of which 9 are associated with I-ICME, 3 with M-ICME, 1 with SIR, and 2 with the intricate interplay between ICME and SIR structures. It is noted that of all the storms in cycle 23, around 23\% are single-peaked with a significant majority of multiple-peaked. However, this trend is precisely not the same in cycle 24, where around 42\% of all the storms are single-peaked.

Further, interestingly, around 23\% and 36\% of single-peaked storms in cycles 23 and 24, respectively, are driven by multiple interacting drivers. Moreover, around 59\% and 67\% of multiple-peaked storms in cycles 23 and 24, respectively, are driven by an individual ICME and SIR structure. It becomes evident that a substantial proportion of multiple-peak storms in both solar cycles, especially in a weaker solar cycle 24, can also trace their origins to isolated ICME/SIR. Multiple steps in Dst may be related to multiple intervals of southward magnetic field separated by less geoeffective conditions in the solar wind driver. This phenomenon can be attributed to the capacity of distinct sub-structures within a single ICME/SIR to inject energy into the Earth's magnetosphere; for example, some ICME-driven storms have separate contributions from the sheath region and the ICME ejecta \citep{Gonzalez1999,Huttunen2002}. Also, a fraction of single-peak storms in both cycles, especially in a weaker cycle 24, can be due to multiple ICMEs and interacting ICME-SIR structures. Based on the storms in both cycles, we emphasize that the number of peaks in Dst is not necessarily directly related to the number of interplanetary drivers associated with a particular storm \citep{Richardson2008a}.

From our analysis of the single-peaked storm of cycle 23, we noted the following Pearson's (Spearman's) correlations between the duration of the recovery phase and various other parameters: -0.66 (-0.74) with the Dst peak, 0.11 (0.23) with the main phase duration, -0.32 (-0.49) with the main phase buildup rate, 0.47 (0.43) with the recovery rate, 0.36 (0.46) with the fast decay phase, and 0.97 (0.94) with the slow decay phase. We consider correlation coefficients ranging from 0.4-0.6 for moderate dependency and 0.6-0.8 for strong dependency. The recovery phase duration strongly depends on the Dst peak and slow decay phase. The total time of the recovery phase only moderately depends on the main phase buildup rate and fast decay phase of the storm. It is clear that the stronger storms with the slow decay phase certainly lead to the prolonged recovery time of the storm. For the multiple-peaked storms of cycle 23, we noted the following Pearson's (Spearman's) correlations between the duration of the recovery phase and various other parameters: -0.22 (-0.39) with the Dst peak, -0.01 (-0.01) with the main phase duration, -0.11 (-0.16) with the main phase buildup rate, 0.58 (0.84) with the recovery rate, 0.72 (0.63) with the fast decay phase, and 0.95 (0.92) with the slow decay phase. It is clear that the duration of the recovery phase of the multiple-peaked storm strongly depends, in order of preference, on the slow decay phase and fast decay phase of the storm, but the peak of Dst and main phase buildup rate do not govern the total duration of the recovery phase. We note that among all the correlations described above, the p-value is much smaller than 0.05 if the estimated correlation coefficient between the two parameters is more than 0.5. However, the p-value is not less than 0.05 for weaker correlations. This shows the statistical significance of our analysis and the validity of our findings.

Figure~\ref{sc23_recov} shows the correlation between the duration of the recovery phase and different characteristics of the single peak (left) and multiple peak storms (right) of solar cycle 23. The identified drivers for each storm are also indicated in the figure. For both the single and multiple-peak storms, the recovery phase strongly depends on the duration of the slow decay phase and moderately on the fast decay phase. However, the fast and slow decay phase strongly governs the recovery phase duration for multiple-peak storms. The duration of the recovery phase does show a moderate dependency on the main phase buildup rate and strong dependency on the Dst peak for single-peaked storms but not for multiple-peaked storms. Since many multiple-peak storms driven by multiple ICMEs in cycle 23 are much more geoeffective than single-peak storms, it is possible that a more potent storm recovery rate depends on the early fast recovery phase in addition to the later slow recovery phase. Earlier studies have suggested that the contribution of tail current and O$^{+}$ charge exchange loss dominate during fast recovery, whereas the contribution of ring current and H$^{+}$ charge exchange loss dominate during late recovery \citep{Feldstein2000,Choraghe2021}. Our study suggests, in general, that the loss mechanisms of magnetospheric current systems for multiple-peak geomagnetic storms could be different than single-peaked geomagnetic storms. We also note that storms driven by SIR or ICME-SIR interaction, in general, have a prolonged recovery time dominated by a later slow recovery phase, for both single and multiple-peak storms.

\subsection{Recovery phase of single-peaked and multiple-peaked storms of solar cycle 24}

In our investigation of the single-peaked solar storm occurring during cycle 24, we analyzed Pearson's (Spearman's) correlations between the duration of the recovery phase and other parameters. We note the following correlations: -0.49 (-0.32) with the Dst peak, 0.15 (0.07) with the main phase duration, -0.47 (-0.52) with the main phase buildup rate, 0.7 (0.85) with the recovery rate, 0.78 (0.78) with the fast decay phase, and 0.88 (0.77) with the slow decay phase. Notably, the duration of the recovery phase for single peak storms strongly depends on the slow and fast decay phase duration, and moderately depends on the main phase buildup rate and Dst peak of the storms.

\begin{figure*}
  
        \centering
        \includegraphics[width=0.42\textwidth]{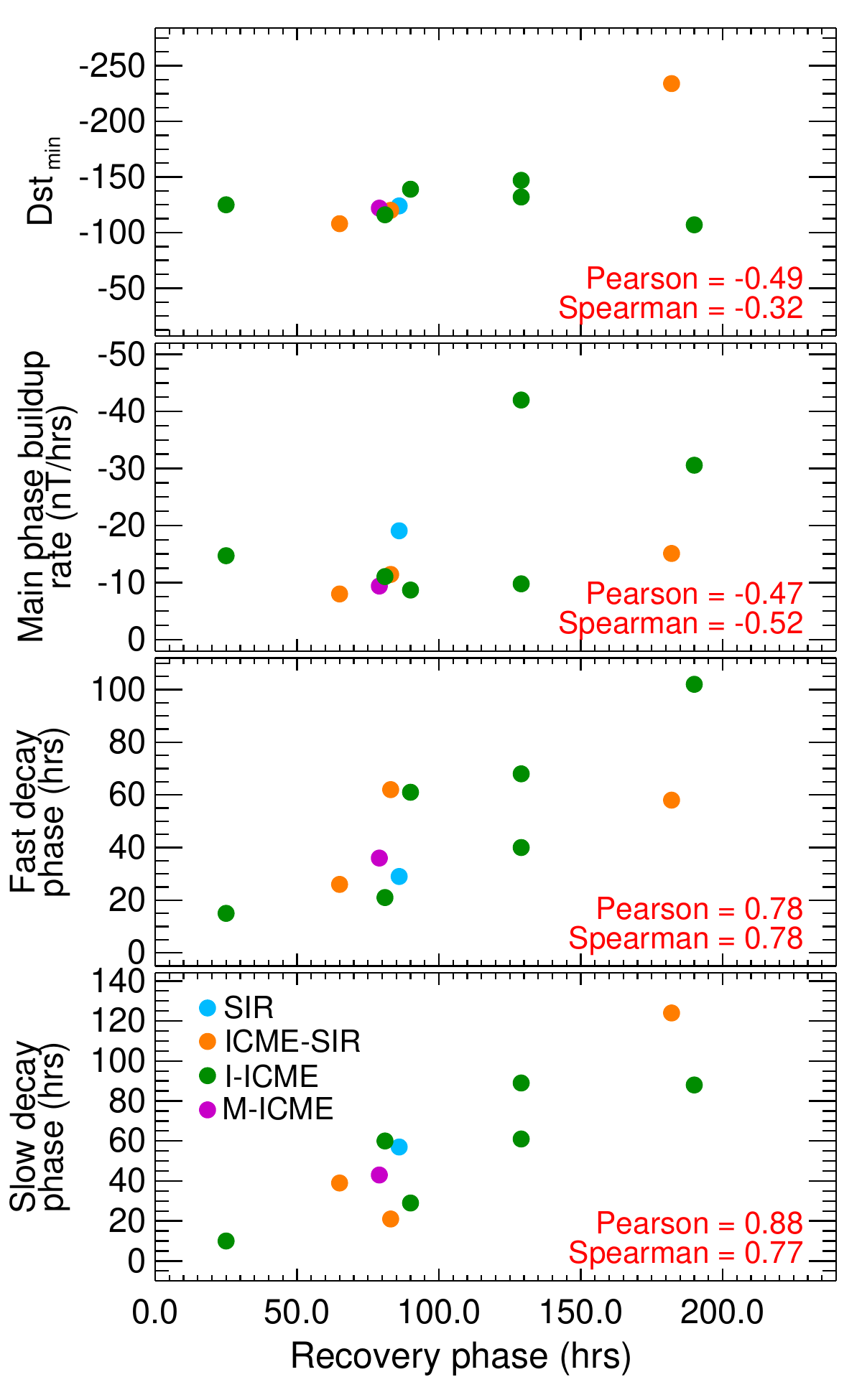}
        \hspace{8mm}
        \includegraphics[width=0.42\textwidth]{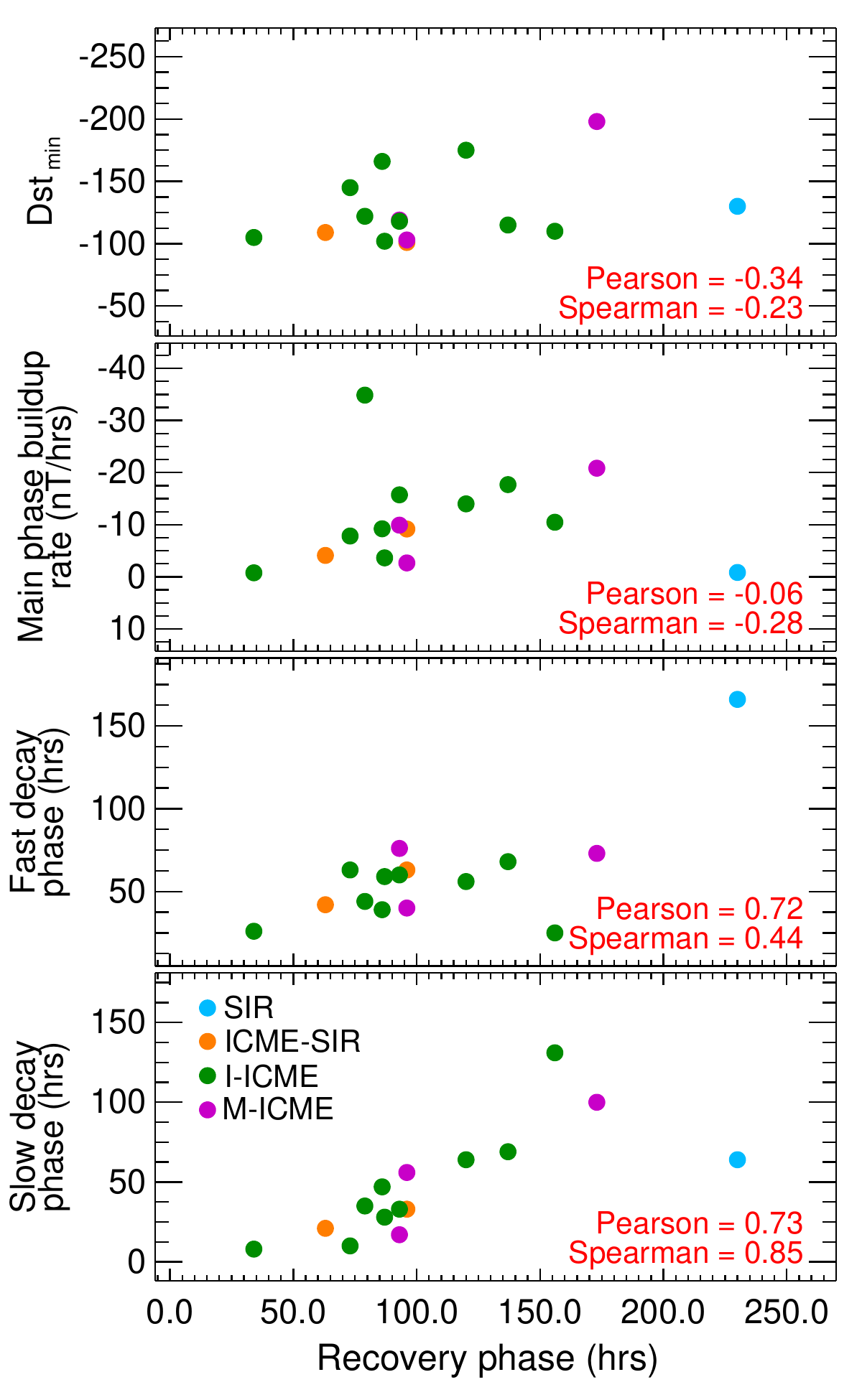}
        
        \caption{Left panel: The correlation between the recovery phase duration of single peak storms of solar cycle 24 and storm characteristics is shown. From the top to the panel, the correlation of the recovery phase with the peak Dst index, main phase buildup rate, duration of the fast decay phase, and slow decay phase, are shown, respectively. Right panel: It is the same as the left panel but for multiple peak storms.}
        \label{sc24_recov}
  
\end{figure*}

In the case of the multiple-peaked storms within cycle 24, we observed the duration of recovery phase has the following correlations: -0.34 (-0.23) with the Dst peak, 0.21 (-0.21) with the main phase duration, -0.05 (-0.28) with the main phase buildup rate, 0.76 (0.87) with the recovery rate, 0.72 (0.44) with the fast decay phase, and 0.73 (0.85) with the slow decay phase. It becomes evident that the recovery phase's duration in multiple-peaked storms strongly depends on the slow and fast decay phase duration. The duration of the recovery phase has no meaningful (not even moderate) dependence on the main phase buildup rate and the Dst peak of the storm. The duration of the recovery phase is significantly influenced by the characteristics of the decay phases (fast and slow) in both single-peaked and multiple-peaked storms in cycle 24. Among the correlations discussed earlier, it is important to highlight that when the estimated correlation coefficient between the two parameters exceeds 0.5, the corresponding p-value is significantly less than 0.05. However, the p-value is not less than 0.05 for weaker correlations. This underscores the statistical significance of our analysis and confirms the validity of our findings.

In Figure~\ref{sc24_recov}, we depict the relationship between the recovery phase's duration and various attributes of both single-peak (on the left) and multiple-peak (on the right) storms during solar cycle 24. The figure also highlights the identified transients driving each of these storms. It is clear that in cycle 24, both the duration of the fast and slow decay phases have a strong correlation with the duration of the recovery phase for the single-peaked and multiple-peaked storms. From the estimated correlation coefficients, we note a non-parametric relation between the recovery phase and duration of the fast decay phase in cycle 24 for multiple-peaked storms.

We note that the duration of the fast decay phase showed only a moderate dependency on the recovery phase duration of single-peaked storms in cycle 23 but a strong dependency in all other cases. The duration of the recovery phase moderately depends on peak Dst for the single-peaked storms of cycle 24, but a strong dependency is noted for single-peak storms of cycle 23. This could seem relatively trivial, but this relation is not seen in multiple-peaked storms, as multiple peaks affect the recovery phase. We acknowledge the limitations of comparing both cycles, as the calculation of correlation coefficients for cycle 24 is from a smaller sample size than cycle 23.

As described in Section~\ref{dstlt100_2324}, we could ascribe the drivers to each storm using the ICMEs and SIRs catalogs \citep{Richardson2010,Shen2017,Chi2018}. Based on a binary classification (i) storms driven by ICMEs (either singular or multiple ICMEs) and (ii) storms driven by SIRs (either from SIRs or ICME-SIR interacting structure), we note that during solar cycle 23, SIR-driven storms exhibited an average main phase duration of approximately 12 hours, while ICME-driven storms had an average main phase duration of about 19 hours. In terms of the recovery phase, SIR-driven storms displayed an average duration of around 145 hours, whereas ICME-driven storms had an average recovery phase duration of 135 hours. Additionally, the average minimum Dst index ($\text{Dst}_{\text{min}}$) for ICME-driven storms during this cycle was -170 nT, compared to -140 nT for SIR-driven storms. In the context of solar cycle 24, the typical duration of the main phase for storms driven by SIRs is larger than those caused by ICMEs. Regarding the recovery phase, SIR-driven storms tend to have an average duration of about 120 hours, whereas ICME-driven storms typically last around 100 hours. Additionally, the average minimum Dst value for storms driven by ICMEs during cycle 24 is -130 nT, while for SIR-driven storms, it registers at -115 nT.

SIRs having a longer recovery phase, for both the cycles, than ICMEs can be attributed to the fact that they are associated with large amplitude Alfvén waves, which are generally found along with high-speed fast streams and, thus, have a fluctuating magnetic field \citep{Echer2013}. Consistent with earlier studies, we also note that ICMEs are more geoeffective than SIRs \citep{Mouikis2019}. In both cycles, a more significant number of strong storms, with Dst$_{\text{min}}$ $\leq-100$ nT, are driven by ICMEs than by SIRs. However, we find a few cases of purely SIR-driven storms going up to $\text{Dst}_{\text{min}}$ values around -150 nT, and thus capable of causing extreme space weather similar to ICMEs. There is clear evidence that solar cycle 24 is significantly weaker than solar cycle 23.

\section{Results and Discussion}

In this work, we have investigated the geomagnetic storm's distribution and recovery phase characteristics for solar cycles 23 and 24. This duration of two cycles included the period from 1996 to 2019. During this period, there were 263 and 148 moderate and stronger-than-moderate geomagnetic storms with Dst$_{\text{min}}\leq -100\;\text{nT}$ in cycles 23 and 24, respectively. The sample included 73 and 26 intense and stronger-than-intense storms with Dst$_{\text{min}}\leq -100\;\text{nT}$ in cycles 23 and 24, respectively, including single and multiple peaks. Our study shows the lack of stronger geomagnetic storms in cycle 24 than in the previous cycle, which confirms earlier study about weaker solar cycle 24 based on various proxies of solar magnetic variability \citep{Kilpua2014,Watari2017,Mishra2019}.

The annual occurrence rate of moderate and stronger-than-moderate geomagnetic storms for both cycles follows the sunspot cycle. However, we note that the two peaks in the rate of such storms do not exactly coincide with those in the sunspot cycle. The first peak in the number of storms coincides with the first peak in the sunspot number, which is close to solar maxima, while the second peak in storms falls a year after the second peak in the sunspot numbers, which is in the early decline phase of the cycle. The annual occurrence rate of intense and stronger-than-intense geomagnetic storms closely follows the peaks in the sunspot cycle 23, but the second peak in the storm's number for cycle 24 falls a year after the second peak in the sunspot cycle. Interestingly, we note that SIR's contribution to intense storms in cycle 23 is 20\%  and 10\%  of total storm numbers in the rise/decline and the maximum of the cycle, respectively. However, for cycle 24, SIR-driven intense storms comprise up to 30\% of the total number of storms in the maximum of the cycle, and it could comprise as much as ICMEs in the rising phase of the cycle. We also note the lack of SIR-driven intense storms towards the beginning and end of cycle 24 during the deep minimum. It is expected that high-latitude CMEs and SIRs from non-sunspot regions give rise to the second peak of the storm's number in both cycles when moderate and stronger-than-moderate storms are included, while it is true only in cycle 24 when intense and stronger-than-intense storms are included \citep{Echer2008}. Therefore, the contribution of SIRs is notable in a weaker solar cycle and while considering moderate storms.

Regarding the monthly distribution of storm occurrence, we note semiannual peaks in storm numbers at equinoxes for cycles 23 and 24. The moderate and stronger-than-moderate storm numbers can reach up to double close to equinoxes (February-April and September-October) than at solstices. However, the prominent peak in cycle 23 is in October, a month after the September equinox, while in February, a month before the March equinox, in cycle 24. The annual variation appears to show maxima near equinoxes, alternating between consecutive cycles as suggested in \citet{Mursula2011}. The monthly distribution of intense and stronger-than-intense storms for cycle 23 shows peaks after two months of equinoxes in May and November, while in March and October for cycle 24. Therefore, we do not find that the dominant peak in intense storms alternates in consecutive cycles, as it comes one or two months after the September equinox in both cycles. The semiannual peaks are closer to equinoxes for moderate storms and weaker cycle 24. It seems that the mechanisms that enhance geoeffectiveness at equinoxes can primarily benefit moderate geomagnetic activities.

Our study focused on understanding the recovery phase duration of the storms and its relation with the storm's other characteristics. Since the recovery phase is strongly affected by the multiple steps in the Dst, we classified the storms of both solar cycles 23 and 24 into two classes: single-peak and multiple-peak storms. We note that of all the intense and stronger-than-intense storms, there are only 23\% storms showing single-peak in cycle 23 while it is 42\% in cycle 24. The study of \citet{Richardson2008a} estimated that around 59\% of the storms might be classified as a single step, while it was estimated as 29\% in the study of \citet{Kamide1998}. Our study finds that a stronger cycle 23 has a significantly larger fraction of multiple-peak storms than a weaker cycle 24. We acknowledge that determining the step count in storms relies on the criteria for identifying multiple peaks, such as strengths and duration of Dst fluctuations. Future studies focused on analyzing the Dst profile in conjunction with geoeffective parameters of solar wind can aid in deducing the presence of these fluctuations or dips.

We further divided the storms into four different classes depending on their identified drivers. These categories are ``I-ICME'' for storms driven by a single ICME, ``M-ICME'' for storms driven by multiple interacting ICMEs, ``SIR'' for those driven by a single SIR, and ``ICME-SIR'' for those driven by interacting ICME and SIR structure. In a stronger cycle 23, multiple ICMEs and ICME-SIR interaction are more probable and likely to give intense storms with multiple peaks than in a weaker cycle 24. However, a significant fraction of multiple-peak storms (~60\% in cycle 23 and ~70\% in 24) can also trace their origins to isolated ICME/SIR, and a smaller fraction of single-peak storms (~25\% in cycle 23 and ~35\% in 24) can be driven by multiple interacting drivers. This has also been suggested based on analyzing a few selected intense storms of solar cycle 23 \citep{Richardson2008a}. Our study analyzing all the storms in both solar cycles confirms that the number of dips in the Dst is not always associated with the number of drivers and vice-versa.  This implies that different steps in storms, close and far-spaced in time, should be analyzed in detail to understand if some magnetospheric processes cause them or simply because of multiple separated intervals of enhanced geoeffective conditions in the solar wind driver.

Our investigation reveals the influence of various storm characteristics on the duration of the recovery phase, as evidenced by the correlation coefficients established among distinct storm parameters. The extended recovery phase observed in single-peak storms within cycle 23 exhibits a strong association with (i) a long-duration slow decay phase and (ii) a notable decline in the Dst peak while demonstrating a moderate correlation with (i) the higher main phase buildup rate (indicative of a more rapid decrease in negative Dst) and (ii) a longer fast decay phase. However, the recovery phase of single peak storms for cycle 24 is strongly related to the duration of both (i) slow decay and (ii) fast decay phases, while moderately related to the (i) main phase buildup rate and (ii) Dst peak. In contrast, for multiple peak storms in cycles 23 and 24, the recovery phase duration is strongly governed by both the slow and fast decay phase duration, while no association is noted with the main phase buildup rate and Dst peak. Interestingly, for both cycles, the recovery phase duration moderately depends on the main phase buildup rate for single-peak storms but not for multiple-peak storms.

The end of the main phase is associated with reversing the polarity of IMF Bz from the southward to northward condition. This signals the onset of the recovery phase when the ring current ceases to receive the supply of energetic ions through the nightside earthward convection. On many occasions, when the IMF Bz suddenly turns northward after being stable and southward for some time, it generates an over-shielding electric field in the entire inner magnetosphere \citep{Chakrabarty2006}. In addition, substorms can also generate an overshielding-like scenario in the inner magnetosphere \citep{Hashimoto2017}. Overshielding events, in general, mark the beginning of the recovery phase of the storm. During substorms, the tail current collapses \citep{Mcpherron1973,Kepko2015}, leading to the reduction in the negative horizontal component measured at the ground. \citet{Huang2004} showed that the Dst/Sym-H index shows similar variations after each substorm onset and proposed that dipolarizations of the nightside magnetic field related to substorm onsets cause these changes. Given the above scenario, it is possible that the fast decay of the recovery phase is associated with the onset of substorms, and hence, relatively poor correlations are found compared to the slow decay (longer duration), particularly in Cycle 23.  The roles of CME and CIR-triggered substorms in the early recovery phase of geomagnetic storms need to be investigated in greater detail to understand these aspects.

The study of \citet{Choraghe2021} suggests that the magnitude of the recovery rate during the slow decay phase is proportional to the Dst peak; however, in our study, it is only true for single-peak storms of cycle 23. 
Our study suggests that the loss mechanism for magnetospheric ring currents responsible for multiple-peak storms could be different than for single-peak storms, and therefore, the origin of multiple peaks needs to be investigated further. We also note that ICMEs, especially interacting ICMEs, are more geoeffective than SIRs, but the SIRs-led storms have a larger recovery time \citep{Chi2018}. It is evident that a stronger geomagnetic storm does not necessarily have a longer main or recovery phase duration, especially for multiple-peak storms, as different current systems are involved at different steps, and their decay rate can vary from storm to storm.

Earlier studies have focused on modeling the complete recovery phase of storms by a mathematical function, and those studies are limited to only a few great or severe storms \citep{Cid2013,Choraghe2021}. In our study, we estimate the main phase and the recovery phase duration of all the storms with $\text{Dst}_{\text{min}}$ $\leq-100\;\text{nT}$ for both solar cycles 23 and 24. Our study also divides the storm's recovery phase into two intervals (early and late recovery) and examines its dependence on its main phase characteristics. This approach is also taken in \cite{Yermolaev2016}, but our finding differs from that reported. \cite{Yermolaev2016} indicated a correlation between storm recovery time and main phase buildup rate in events induced by CIR and sheath compression regions, but not in those induced ICMEs. Our study has classified the storms as single or multiple-peaked storms to understand the recovery phase dependent on different drivers.


Our investigation into the recovery phase of storms induced by various drivers is confined to intense and stronger-than-intense storms ($\leq-100\;\text{nT}$). Consequently, the findings we have drawn may not be directly applicable to moderate storms. It would be interesting to explore the recovery phase characteristics of moderate storms in future research to determine if they exhibit similar patterns. We acknowledge that categorizing storms into single and multiple-steps, using a specific criterion, oversimplifies their complexity. Nevertheless, our study analyses the annual and solar cycle distribution of all storm events, encompassing those in the moderate category. In our analysis, we have identified the driving factors for each storm from distinct catalogs \citep{Richardson2010, Shen2017, Chi2018}, without distinguishing between the sheath, leading, or trailing portions of the ejecta as potential drivers for the storms. Our study does not consider whether interacting structures arrive at Earth as individual entities or complex ejecta \citep{Burlaga2002,farrugia2004,Lugaz2014,Mishra2014}. Such differentiation may offer insights into the multiple peaks observed in storm profiles and their subsequent recovery phases. Our future research plans include delving deeper into this direction and conducting a detailed analysis of in-situ solar wind observations near Earth. Furthermore, a comprehensive and in-depth investigation into modeling the development and decay of the Dst index for storms driven by diverse factors is warranted.


\section*{Acknowledgements}

We thank the ACE, WIND team members, and different geomagnetic observatories for making their observations publicly available. We also acknowledge using the ICME catalog compiled by Richardson and Cane and the SIR catalog compiled by Yutian Chi. The authors thank the referees for their constructive comments on improving the manuscript. 

\section*{Data Availability}

The data sets used and generated for this study are available on request to the corresponding author.






\appendix

\section{Some extra material}

If you want to present additional material which would interrupt the flow of the main paper, it can be placed in an Appendix which appears after the list of references.


\bsp	
\label{lastpage}
\end{document}